\documentclass{aa}  
\usepackage{amssymb}
\usepackage{graphicx}
\usepackage[authoryear]{natbib}
\usepackage[figuresright]{rotating}

\usepackage{txfonts}
\newcommand{\heia}{\ion{He}{i} 2.058 $\mu m$}
\newcommand{\heib}{\ion{He}{i} 2.112 $\mu m$}

\newcommand{\heiia}{\ion{He}{ii} 2.189 $\mu m$}
\newcommand{\heiib}{\ion{He}{ii} 2.037 $\mu m$}
\newcommand{\heiic}{\ion{He}{ii} 2.346 $\mu m$}
\newcommand{\brg}{Br$\gamma$}

\newcommand{\niiia}{\ion{N}{iii} 2.247, 2.251 $\mu m$}

\newcommand{\civa}{\ion{C}{iv} 2.070-2.084 $\mu m$}
\newcommand{\siiv}{\ion{Si}{iv} 2.428 $\mu m$}
\newcommand{\mum}{\ifmmode \mu \rm{m} \else $\mu \rm{m}$\fi}

\newcommand{\teff}{\ifmmode T_{\rm eff} \else T$_{\mathrm{eff}}$\fi}
\newcommand{\logg}{\ifmmode \log g \else $\log g$\fi}
\newcommand{\lL}{\ifmmode \log \frac{L}{L_{\odot}} \else $\log \frac{L}{L_{\odot}}$\fi}
\newcommand{\mdot}{$\dot{M}$}
\newcommand{\myr}{M$_{\odot}$ yr$^{-1}$}

\newcommand{\vinf}{$v_{\infty}$}
\newcommand{\vturb}{v$_{\rm turb}$}
\newcommand{\vesc}{v$_{esc}$}
\newcommand{\kms}{km s$^{-1}$}
\newcommand{\msun}{\ifmmode M_{\odot} \else M$_{\odot}$\fi}
\newcommand{\zsun}{\ifmmode Z_{\odot} \else Z$_{\odot}$\fi}
\newcommand{\lsun}{\ifmmode L_{\odot} \else L$_{\odot}$\fi}
\newcommand{\rsun}{\ifmmode R_{\odot} \else R$_{\odot}$\fi}
\newcommand{\qh}{\ifmmode Q_{\rm H} \else $Q_{\rm H}$\fi}
\newcommand{\qhei}{\ifmmode Q_{\ion{He}{i}} \else $Q_{\ion{He}{i}}$\fi}

\begin{document}
   \title{The most massive stars in the Arches cluster \thanks{Based on observations collected at the ESO Very Large Telescope (program 075.D-0736(A))}}

   \subtitle{}

   \author{F. Martins\inst{1}
          \and
          D.J. Hillier\inst{2}
          \and
          T. Paumard\inst{3}
          \and
          F. Eisenhauer\inst{1}
          \and
          T. Ott\inst{1}
          \and
          R. Genzel\inst{1,4}
          }

   \offprints{F. Martins}

   \institute{Max-Planck Instit$\ddot{\rm u}$t f$\ddot{\rm u}$r extraterrestrische Physik, Postfach-
1312, D-85741, Garching bei M$\ddot{\rm u}$nchen, Germany \\
              \email{martins@mpe.mpg.de}
         \and
             Department of Physics and Astronomy, University of Pittsburgh, 3941 O'Hara St., Pittsburgh, PA 15260, USA\\
	 \and
             LESIA, Observatoire de Paris, CNRS, UPMC, Universit\'e Paris Direrot, 5 Place Jules Janssen, F-92195, Meudon CEDEX, France\\
         \and
             Department of Physics, University of California, CA 94720, Berkeley, USA\\
             }

   \date{Received 13 August 2007; accepted 5 November 2007}

 
  \abstract
   {}
   {We study a sample composed of 28 of the brightest stars in the Arches cluster. Our aim is to constrain their stellar and wind properties and to establish their nature and evolutionary status.}
   {We analyze K-band spectra obtained with the integral field spectrograph SINFONI on the VLT. Atmosphere models computed with the code CMFGEN are used to derive the effective temperatures, luminosities, stellar abundances, mass loss rates and wind terminal velocities.}
   {We find that the stars in our sample are either H-rich WN7--9 stars (WN7--9h) or supergiants, two being classified as OIf$^{+}$. All stars are 2--4 Myr old. There is marginal evidence for a younger age among the most massive stars. The WN7--9h stars reach luminosities as large as $2 \times 10^{6} L_{\odot}$, consistent with initial masses of $\sim$ 120 \msun. They are still quite H-rich, but show both N enhancement and C depletion. They are thus identified as core H-burning objects showing products of the CNO equilibrium at their surface. Their progenitors are most likely supergiants of spectral types earlier than O4--6 and initial masses $>$ 60 \msun. Their winds follow a well defined modified wind momentum -- luminosity relation (WLR): this is a strong indication that they are radiatively driven. Stellar abundances tend to favor a slightly super solar metallicity, at least for the lightest metals. We note however that the evolutionary models seem to under-predict the degree of N enrichment. }
   {}


   \maketitle


\section{Introduction}
\label{s_intro}

The center of our Galaxy is a unique environment to study massive
stars. It harbors three massive clusters -- the Arches, Quintuplet and
central cluster -- which together contain about 30\% of the number of
Wolf-Rayet stars known in the Galaxy \citep{vdh06}. Interestingly, the
three clusters have different ages, ranging from $\sim$ 2 Myr for the
Arches to $\sim$ 6 Myr for the central cluster. Consequently, they
host different populations of massive stars and sample the entire
upper part of the HR diagram. Studying their stellar content gives us
a unique opportunity to understand how massive stars evolve.

Although there is a global framework explaining the evolution of stars
more massive than $\gtrsim$ 20 \msun, a quantitative description is
still lacking. According to \citet{paul95}, stars with masses in the
range 25--60 \msun\ experience the sequence O $\rightarrow$ Of
$\rightarrow$ LBV or RSG $\rightarrow$ WN8 $\rightarrow$ WNE
$\rightarrow$ WC, while for more massive stars, the sequence O
$\rightarrow$ Of $\rightarrow$ WNL+abs $\rightarrow$ WN7
($\rightarrow$ WNE) $\rightarrow$ WC is preferred. \citet{langer94}
favor another scenario in which all stars have a H-rich WN phase prior
to a LBV event: O $\rightarrow$ H-rich WN $\rightarrow$ LBV
$\rightarrow$ H-poor WN $\rightarrow$ H-free WN $\rightarrow$ WC.  We
see that there are still some qualitative differences between the
proposed scenarios. Further, the evolutionary sequences are not mature
enough to allow a refinement of the classification of the stars in the
different evolutionary states. For example, the spectral types of the
O or WC stars entering the above scenarios are not specified. The
question of whether or not all massive stars go through a LBV phase is
also not answered. This is an important caveat, especially since this
phase has recently been argued to be the one in which most of the mass
removal happens \citep{so06}.

In a previous study \citep{gc07}, we analyzed 18 massive stars in the
central cluster of the Galaxy. This cluster is especially intriguing
since it hosts the supermassive black hole SgrA*
\citep{genzel96,ghez98}. In spite of the drastic tidal forces, several
tens of massive stars formed recently
\citep{ahh90,krabbe95,pgm06}. Some of them are approaching the black
hole at distances of only a few light hours
\citep{ghez03,frank05}. The presence of young stars in the Galactic
Center together with the apparent implausibility of forming stars so
close to the central supermassive black hole is a puzzle usually
referred to as ``the paradox of youth''. Studying the dynamics of
these young stars, \citet{pgm06} \citep[see also][]{lb03,genzel03}
have shown that they orbit SgrA* in two counter-rotating
disks. Together with other evidences (total mass and structure of the
disks), this points to a local, ``in-situ'' star formation event. The
detailed analysis of the post-main sequence massive stars has revealed
that, surprisingly, their evolution follows almost perfectly the
predictions of evolutionary models \citep{gc07}. This implies that
whatever the exact formation mechanism is, the subsequent evolution is
not different from that predicted for normal stars. We found that all
stars seem to have progenitors in the mass range 25--60 \msun\ and
that they follow relatively well the evolutionary scenario proposed by
\citet{paul95} for this mass range. We have been able to refine this
scenario, pinpointing the relation between different spectral types: O
$\rightarrow$ (Ofpe/WN9 $\leftrightarrow$ LBV) $\rightarrow$ WN8
$\rightarrow$ WN8/WC9 $\rightarrow$ WC9. This was made possible by the
detailed study of stellar abundances in various Wolf-Rayet stars and
related objects. Abundance analysis is a powerful tool to constrain
stellar evolution since it gives direct access to the evolutionary
state of a star.

The above study focused on stars in the mass range 25--60 \msun\ due
to the age of the central cluster (more massive stars do not exist any
more). In order to constrain stellar evolution at very high mass, we
need to study younger clusters. The Arches cluster in the Galactic
Center, only 30 pc away from the central cluster, is the ideal
target. Not only is it believed to be quite young (2 to 4.5 Myr, see
Figer et al.\ 1999, Blum et al. 2001 and the present study), but it
also shares the same environment as the central cluster, and hence
have the same metallicity. Hence, its study ensures to obtain a
homogeneous view of stellar evolution among all types of massive stars
in the Galactic Center. The Arches cluster, first discovered by
\citet{nagata95} and \citet{cotera96}, is also one the most massive
and densest cluster of the Galaxy. \citet{figer99} first showed that
the mass function (MF) of its central regions might be shallower than
the typical Salpeter IMF. This result was confirmed by
\citet{stolte02}. Although there are indication that high mass star
formation might be favored in the Galactic Center
\citep{ms96,klessen07}, recent simulations of the dynamical evolution
of the cluster by \citet{kim06} indicate that we might in fact witness
the effects of mass segregation in the cluster core rather than an
actual top-heavy initial mass function. Whatever the physical reason,
there are nearly 100 massive stars in the Arches cluster. From K-band
spectroscopy of the brightest members, \citet{blum01} and
\citet{figer02} identified several late WN stars and early O
supergiants. The analysis of five of these stars by \citet{paco04}
revealed, by an indirect method, that their metallicity was close to
solar.

Here, we analyze a much larger sample (28 stars in total) in order to
better constrain their stellar and wind parameters. We rely on high
quality data obtained with the integral field spectrograph SINFONI on
the VLT. The K-band spectra extracted from this data set are analyzed
with atmosphere models computed with the code CMFGEN \citet{hm98}. In
Sect.\ \ref{s_spec} we describe the observations, our sample and the
spectroscopic classification; in Sect.\ \ref{s_analysis} our method to
analyze the stars are presented; the results are summarized in Sect.\
\ref{s_res} and discussed in Sects.\ \ref{s_hr}, \ref{s_nature_wn7}
and \ref{s_Z}. We give our conclusions in Sect.\ \ref{s_conc}.


\section{Spectroscopic data}
\label{s_spec}

\subsection{Observations and data reduction}
\label{s_obs}

The Arches cluster was observed in service mode between May 3$^{\rm
rd}$ and June 27$^{\rm th}$ 2005 with SINFONI on the ESO/VLT
\citep{spiffi,bonnet04}. K band data were obtained with a pixel scale
of 0.1\arcsec. Adaptive optics was used to improve the spatial
resolution. The seeing varied between 0.5 and 1.2 arcseconds during
the different runs. Four sub-fields were observed at the core of the
cluster, as well as 12 fields in the outer part. For each sub-field,
the integration time on source was 240 seconds. One sky exposure was
obtained every two object exposures. Early B stars were observed as
telluric standards.

Data were reduced with the SPRED software \citep{spred} as in
\citet{frank05} and \citet{pgm06}. The reduction steps include: sky
subtraction, flat field and bad pixel correction, distortion
correction, wavelength calibration and atmosphere correction. In the
last step, the telluric standard is used after its intrinsic \brg\
line is removed by a simple interpolation of the continuum red and
blueward of the line. Individual frames were subsequently combined to
obtain mosaics of the observed regions (when frames overlap). We refer
the reader to \citet{spred} for a comprehensive description of the
software used. Spectra were then carefully extracted by selection of
individual ``source'' pixels from which ``background'' pixels are
removed to correct for light contamination. The final spectra have a
resolution of $\sim$ 4000 and a signal to noise ratio of 10 to 80
depending on the brightness of the star.

\subsection{Sample }
\label{s_sample}

We selected the stars with high enough signal to noise ratio spectra
(S/N $\gtrsim$ 10) for a subsequent quantitative analysis with
atmosphere models. Equivalently, this means that we studied the
brightest members of the Arches cluster. The list is given in Table
\ref{tab_phot}. The name of the stars is taken from the list of
\citet{figer02}. NICMOS photometry in the F205W filter was taken from
\citet{figer02}, and was assumed to be equivalent to K-band
photometry. We also included star number 1 of \citet{blum01} which is
not in the list of \citet{figer02}. It is designated by the name
B1. Its K-band magnitude is taken as the 2.14\mum\ magnitude of
\citet{blum01}. This wavelength range is free of strong line. We
estimated the absolute K-band magnitudes adopting a distance to the
Galactic Center of 7.62 kpc \citep{frank05}. We also adopted a
constant extinction $A_{\rm K} = 2.8$ for all stars. This value is
slightly lower than the average $A_{\rm K}$ derived by
\citet{stolte02} and \citet{kim06}. However, as noticed by these two
studies, the extinction is smaller in the cluster center (inner
5\arcsec). In this region, $A_{\rm K}$ is between 2.6 and 2.95
\citet{stolte02}. This behavior is interpreted as evidence for swiping
of dust by stellar winds and/or photo-evaporation by the intense UV
radiation of massive stars. Since most of the stars in our selected
sample are in the cluster center, the choice of $A_{\rm K} = 2.8$ is a
reasonable assumption. From the adopted distance and extinction, we
can derive the absolute magnitudes of our sample stars: they are
reported in column 3 of Table \ref{tab_phot}.

\subsection{Spectral classification}
\label{s_classif}

Spectral classification in the K band is more difficult than in the
classical optical range due to the limited number of lines. However,
catalogs of K band spectra of objects with spectral types derived from
optical studies are becoming available, making the spectral
classification easier \citep{morris96,hanson96,figer97,hanson05}. The
main lines observed in our SINFONI spectra are the following: \heia,
\heiia, \heiib, \heiic, \brg, \niiia, \civa, \siiv\ and the complex at
2.112-2.115 \mum\ (composed of \ion{He}{i}, \ion{N}{iii}, \ion{C}{iii}
and \ion{O}{iii}). They are identified in Fig.\ \ref{fit_all_1} to
\ref{fit_all_4}. The most prominent line, \brg, goes from a strong
emission in the brightest stars to an absorption profile when the
stars become fainter. \heiia\ shows a similar behavior. \heiib\ and
\heiic\ have a weak P-Cygni profile when present. \civa, \niiia\ and
\siiv, when present, are always in emission. Finally, the 2.112--2.115
\mum\ complex is in emission in most of the spectra.

The K band spectra we obtained are typical of late WN (WNL) and early
O type stars, in agreement with \citet{figer02}. WN stars earlier than
WN7 have \heiia\ stronger than \brg, which is observed in none of our
sample stars. The distinction between late WN subclasses is difficult
when only the K band is available \citep{morris96}. In WN7 stars, the
\heiia\ emission is strong (although not as much as \brg). In later
type stars, \heiia\ is much weaker. In general, this morphology is
associated with a strong \heia\ emission. However, in H-rich late type
WN stars (the so-called WNh stars) \heia\ can be seen in absorption
\citep{paul96,cb97,bc99}, which can be qualitatively understood by a
lower He content.  According to the atlas of \citet{hanson96} and
\citet{hanson05}, late type O stars have a weak \heiia\ absorption,
and the 2.112-2.115 \mum\ line complex in absorption or weak
P-Cygni. Our sample stars do not contain these spectral
morphologies. We thus have only early type O stars (O4--6). Among
them, main sequence stars can be distinguished from supergiants by the
shape of \brg: it is in absorption on the main sequence and either
absent (because filled by wind emission) or in emission (usually with
a weak central absorption) in supergiants. Some O supergiants have
stronger lines than standard O4--6I stars (especially \brg) and are
identified as OIf$^{+}$ supergiants\footnote{Historically, the index f
denotes stars with strong \ion{N}{iii} 4634-4641 \AA\ and \ion{He}{ii}
4686 \AA\ emission, while the symbol ``+'' refers to stars showing
\ion{Si}{iv} 4089-4116 \AA\ in emission.}. \citet{figer02} argued that
the distinction between WNL (especially WN7) and OIf$^{+}$ is very
difficult. Both types have strong emission lines, but WNL stars
usually have \heiia\ in emission or at least with a P-Cygni profile
while OIf$^{+}$ have mainly \heiia\ in absorption. However,
\citet{conti95} showed that exceptions exist: HD16691 and HD190429 are
two OIf$^{+}$ with \heiia\ in emission. \citet{figer02} suggested that
the presence of the \niiia\ emission in WNL stars but not in OIf$^{+}$
supergiants could be used to break the degeneracy in the spectroscopic
classification. Based on these considerations, we classify the Arches
SINFONI spectra using the following criteria (restricted to early O
and late WN stars):

\begin{itemize}

\item[$\bullet$] WNL stars have a strong \brg\ and 2.112--2.115 \mum\ emission. \brg\ is stronger than the 2.112--2.115 \mum\ complex. They show \niiia\ (emission). \heia\ is weak and/or in absorption, so all our WNL stars are H-rich (which is confirmed by our quantitative analysis, see below) and thus are classified as WNh. When \heiia\ is purely in emission the spectral type WN7--8h is assigned. When it is in absorption or shows a P-Cygni profile, the star is classified as WN8--9h.

\item[$\bullet$] OIf$^{+}$ have \heiia\ in absorption, a weak or no \niiia\ line, the 2.112--2.115 \mum\ line complex and \brg\ in emission. The strength of \brg\ is similar to that of the 2.112--2.115 \mum\ complex.

\item[$\bullet$] O supergiants (OI) have the same morphological properties as OIf$^{+}$ stars, except that \brg\ is in emission but weaker than the 2.112--2.115 \mum\ complex, or in weak absorption. 

\end{itemize}

\noindent These criteria remain qualitative on purpose, so that broad
groups of stars can be defined without preventing the possibility that
some stars are intermediate between the groups, as is likely to be the
case in a population of massive stars with such a narrow age
spread. The criteria we defined should also be seen as
\textit{relative} criteria to compare the different stars of our
sample. Our final spectral classification for each star is given in
Table \ref{tab_phot}. The spectra of our sample stars are also
displayed in Fig.\ \ref{fit_all_1}, \ref{fit_all_2}, \ref{fit_all_3}
and \ref{fit_all_4}.

\begin{table}
\begin{center}
\caption{Photometry of the stars analyzed in the present paper. The stars are identified by their number in the list of \citet{figer02}. Observed magnitudes are also from this source. A distance of 7.62kpc is assumed \citep{frank05}, as well as a uniform extinction A$_{\rm K}=2.8$ in the K band \citep{stolte02}.  \label{tab_phot}}
\begin{tabular}{lrrr}
\hline\hline
Star   &  ST  &  m$_{\rm K}$  &  M$_{\rm K}$ \\
\hline 
B1 & WN8--9h        &   11.11        &  -6.10 \\
F1  & WN8--9h       &   10.45        &  -6.76 \\
F2  & WN8--9h       &   11.18        &  -6.03 \\
F3  & WN8--9h       &   10.46        &  -6.75 \\
F4  & WN7--8h       &   10.37        &  -6.84 \\
F5  & WN8--9h       &   10.86        &  -6.35 \\
F6  & WN8--9h       &   10.37        &  -6.84 \\
F7  & WN8--9h       &   10.48        &  -6.73 \\
F8  & WN8--9h       &   10.76        &  -6.45 \\
F9  & WN 8--9h      &   10.77        &  -6.44 \\
F10 & O4--6If$^{+}$&   11.46        &  -5.75 \\
F12 & WN7--8h       &   10.99        &  -6.22 \\
F14 & WN8--9h       &   11.22        &  -5.99 \\
F15 & O4--6If$^{+}$&   11.27        &  -5.94 \\
F16 & WN8--9h       &   11.40        &  -5.81 \\
F18 & O4--6I       &   11.63        &  -5.58 \\
F20 & O4--6I       &   12.16        &  -5.05 \\
F21 & O4--6I       &   11.77        &  -5.44 \\
F22 & O4--6I       &   12.02        &  -5.19 \\
F23 & O4--6I       &   12.19        &  -5.02 \\
F26 & O4--6I       &   12.34        &  -4.87 \\
F28 & O4--6I       &   12.17        &  -5.04 \\
F29 & O4--6I       &   12.26        &  -4.95 \\
F32 & O4--6I       &   12.42        &  -4.79 \\
F33 & O4--6I       &   12.42        &  -4.79 \\
F34 & O4--6I       &   12.49        &  -4.72 \\
F35 & O4--6I       &   12.37        &  -4.84 \\
F40 & O4--6I       &   12.67        &  -4.54 \\
\hline
\end{tabular}
\end{center}
\end{table}


\section{Spectroscopic modeling of individual stars}
\label{s_analysis}

In this section we describe our method to derive the stellar and wind
properties of the selected stars. We also present an estimate of the
uncertainties on the derived parameters.

\subsection{Atmosphere models}
\label{s_code}

We used the atmosphere code CMFGEN \citep{hm98} to derive the stellar
and wind properties of a sample of Wolf-Rayet and O stars. CMFGEN
calculates non-LTE atmosphere models with winds and includes a robust
treatment of line-blanketing. A detailed description of the code was
given by \citet{hm98}. Its main characteristics are also presented in
\citet{n81,ww05,gc07}. Here, we simply highlight a few important
features:

\begin{itemize}

\item[$\diamond$] \textit{hydrodynamic structure}: CMFGEN does not
compute self-consistently the density (and velocity) structure of the
atmosphere. The standard procedure consists in adopting a
pseudo-hydrostatic structure on top of which a ``$\beta$ velocity
law'' is connected. Such a law is expected from theoretical ground
\citep[e.g.][]{pauldrach86}. Here, we used the TLUSTY structures of
\citet{lh03} for the hydrostatic part of the atmosphere. We chose
$\beta = 0.8$ since it is the typical value for O stars
\citep{puls96,repolust04}. It also leads to good fits of the of the
observed line profile. Only for stars B1 and F5 did we have to use
values of 1.2 and 1.8 respectively to better fit the overall shape of
the emission lines, especially \brg: larger $\beta$ produce more
centrally peaked lines (for the adopted clumping factor, see
below). High values of $\beta$ are also found for O supergiants
\citep{paul02,hil03}. The use of such a structure may be questionable
in the case of extreme supergiants and Wolf-Rayet stars. However, the
wind density in these stars is usually so large that the photosphere
is beyond the hydrostatic layers, so that the underlying structure has
little impact on the observed spectrum.

\item[$\diamond$] \textit{line-blanketing and super-levels}: CMFGEN
includes line-blanketing through the super-level approximation. In
practice, levels of similar energies are grouped in a single
super-level which is then used to compute the atmospheric
structure. Within a super-level, individual levels have the same
departure coefficient from LTE. This is a very convenient way to treat
directly line-blanketing, without using statistical methods such as
opacity sampling. Usually, only levels with large energies are grouped
into super-levels. This may affect the strength of infrared lines
since they mainly arise from transitions between such high energy
levels. As a consequence, we decided not to use super-levels for H,
He, NIII and CIV which contribute most of the K band lines: we used
the full atom. In addition to H, He, C and N, we included O, Si, S and
Fe in our models. No other elements could be treated due to the
increased memory requirement implied by the large number of levels
(typically about 3000 levels and 1300 super-levels). The elements
included are anyway the ones responsible for most of the
line-blanketing effects. Tests with additional species confirmed that
adding other elements did not significantly affect the
results. Similarly, tests run with different super-levels assignments
lead to minor changes insufficient to modify quantitatively our
results.

\item[$\diamond$] \textit{microturbulent velocity}: a value of 15
\kms\ was used in the computation of the atmospheric structure. For
the detailed emergent spectrum resulting from the formal solution of
the radiative transfer equation, we adopted \vturb\ = 10 \kms. This is
a reasonable value for O stars \citep{vh00}. For WR stars, it might be
a little too low. However, test models with \vturb\ = 50 \kms indicate
barely any change in the resulting line profiles, because they are
dominated by the wind.

\item[$\diamond$] \textit{clumping}: CMFGEN allows the inclusion of
clumping. In practice, a volume filling factor approach is used. The
filling factor is described by an exponential law starting from a
value of 1.0 at the base of the wind and declining to $f$ in the outer
atmosphere, where the velocity reaches \vinf. Given the limited number
of diagnostics in the K band, we have adopted $f = 0.1$ in our
computations. This is a standard value for WR stars
\citep{hk98,hm99,morris00,hil01,paul02,paul06}. For O stars, the
amount of clumping is still a matter of debate. Values of $f$ as low
as 0.01 have been derived by \citet{jc05,ww05,fullerton06}. But larger
values are also found: $f=0.2$ by \citet{repolust04} or $f=0.1$ by
\citet{paul02b,ww05}. Recently, \citet{puls06} also showed that the
clumping factor might vary non monotonically with radius in the wind,
contrary to the usual assumption of atmosphere models. Given these
uncertainties, the adopted value for $f$ (0.1) is not unrealistic. If
the clumping factor was smaller, then our mass loss rates for O
supergiants would be overestimated by $\sqrt{0.1/f}$.

\end{itemize}

\subsection{Method}
\label{s_method}

We briefly describe here the method we used to constrain the main stellar and wind parameters.

\begin{itemize}

\item[$\diamond$] \textit{Effective temperature}: \teff\ was constrained from the strength of \ion{He}{i} and \ion{He}{ii} lines, as in most studies of massive stars. In practice, we used \heia, \heib, \heiia, \heiib\ and \heiic\ as the main diagnostics. We note that \teff\ is defined at the radius in the atmosphere model where the Rosseland optical depth reaches 2/3. For comparison with evolutionary models, it is usually useful to define T$^{*}$ as the temperature where the opacity is 20. This corresponds to a deeper, quasi hydrostatic layer of the atmosphere, which is more similar to the outer radius of the evolutionary models. In general, \teff\ and T$^{*}$ are almost identical for O stars, while for WR stars with denser winds they can differ by several thousands of degrees.

\item[$\diamond$] \textit{Luminosity}: the K-band flux was used as the main indicator. The luminosity was adjusted so that the K-band magnitude of the models could match the absolute magnitude of the stars. In practice, the K-band flux depends not only on the stellar luminosity, but also on the mass loss rate since in the near infrared the ionized wind produces free-free emission. Hence, the luminosity was derived in parallel to \mdot.

\item[$\diamond$] \textit{Mass loss rate and He abundance}: \mdot\ and the ratio of H to He content were constrained from the strength of the emission of \brg\ and the He lines. A change of \mdot\ leads to a general increase of the emission in all the lines, while an increase of He/H strengthens the He lines and weakens \brg. 

\item[$\diamond$] \textit{C and N abundances}: The carbon and nitrogen content was derived from \civa\ and \niiia\ respectively. We also use the \siiv\ line when observed to constrain the Si abundance. We note however that this last abundance determination is less reliable than for the other elements since the spectrum is much noisier at the position of the \siiv\ line (at the red end of the K band).

\item[$\diamond$] \textit{Terminal velocity}: the terminal velocity of the wind (\vinf) was determined from the width of the \brg\ and the extend of the absorption dip of the P-Cygni profile of \heia\ (when present). When none of these indicators could be used, we simply adopted $v_{\infty} = 2.6 \times v_{\rm esc}$ \citep{lamers95}, $v_{\rm esc}$ being the escape velocity. 

\end{itemize}

\noindent Due to the absence of strong gravity indicators, we adopted \logg\ = 3.25 for the coolest stars, and \logg\ = 3.50 for the hottest. This is a reasonable assumption in view of the calibrations of \citet{msh05}. Finally, the solar abundances of \citet{gs98} were used as references\footnote{The solar abundances have been recently revised for the lightest elements \citep{asplund}, but not for the heaviest. Hence, we prefer to stick to the old values.}.

\subsection{Accuracy of parameters determination}
\label{s_acc}

The determination of the stellar and wind parameters is a long
iterative process: most spectral diagnostics depend on several
parameters. To estimate the uncertainties on our determinations, one
would ideally need to run tens of models covering the parameter space
from which some kind of chi-square procedure (to be defined) could
provide statistical errors. In practice, this is not possible in the
present approach since this would lead to a prohibitively long
process: a model and the associated spectrum require between 24 and 48
hours of cpu time; sampling correctly the parameter space (10 to 20
models for each parameter, and $\sim$ 10 parameters) for each of the
28 stars would imply several months of computations. This is in
addition to the time needed to actually find the best fit model for
each star.

\begin{figure}
\includegraphics[width=9cm]{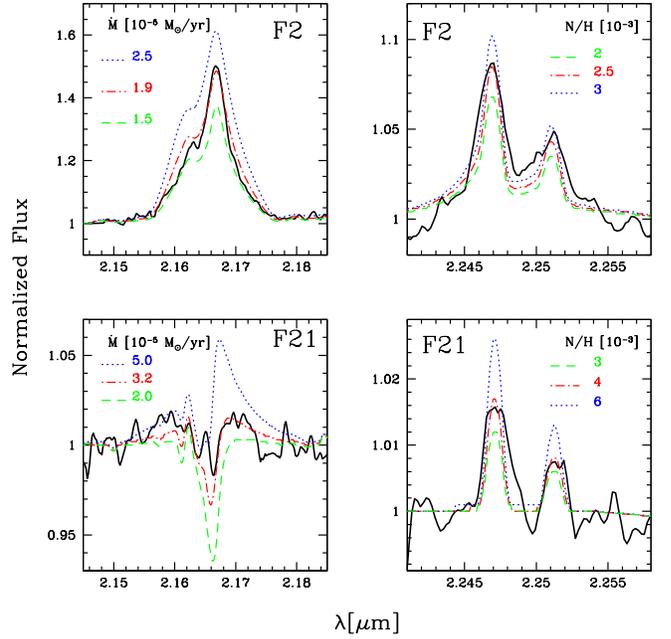}
\caption{Uncertainty estimate for \mdot\ and N/H for one WNLh star (F2) and one O supergiant (F21). The solid line is the observed spectrum, while the colored broken lines are the models. The red dot-dashed lines correspond to the best fit model. In the other models, only \mdot\ (left panels) or N/H (right panels) have been varied. \label{error_param}}
\end{figure}

Hence, we prefer to rely on a more empirical way to estimate the
uncertainties. For this, we chose to run a few test models for two
typical stars (one WN8--9h and one O supergiant). We varied the
parameters around the values of the best fit model and judged by eye
when the resulting spectrum was not satisfactory any more: this was
used to define our uncertainty. This gives a reasonable estimate of
the accuracy with which the stellar and wind parameters are
derived. Fig.\ \ref{error_param} illustrates this procedure
for \mdot\ and N/H.

In practice, we focused on the models for star F2 (WN8--9h) and star
F21 (O4--6I). They were chosen as typical of their class of objects,
both in terms of spectral morphology and derived parameters. The
typical errors are: $\pm$3000K (2000K) on \teff\ for O stars (WNLh),
$\pm$0.2 dex on \lL, 0.2 (0.1) $\log \dot{M}$, 100 \kms\ on terminal
velocities and $\pm$50\%\ ($\pm$30\%) on abundances. The uncertainty
on \lL\ depends mainly on the adopted distance and extinction, and is
thus similar for WNLh and O stars. The uncertainty on abundances does
not take into account any possible systematics due to uncertainties in
atomic data and model assumptions.

\section{Derived stellar and wind parameters}
\label{s_res}

The results of our analysis are summarized in Table \ref{tab_res}. The
final fits are shown in Figs.\ \ref{fit_all_1}, \ref{fit_all_2},
\ref{fit_all_3} and \ref{fit_all_4}. A summary of the general
properties is given in Sect.\ \ref{s_general} while a comparison to
previous studies is made in Sect.\ \ref{s_comp_paco}.

\subsection{General properties}
\label{s_general}

To avoid lengthy discussions, we do not describe the stars one by one but focus on their global properties. They can be summarized as follows:

\begin{itemize}

\item[$\diamond$] \textit{The stars have very similar \teff}: out of 28 stars, 17 have $32500 < \teff < 37500$. The average effective temperatures of WN7--9h and O4--6I stars are 33600 K and 37600K respectively. Considering the error on the temperature estimate ($\pm$2000--3000 K), there is very little dispersion in \teff\ among our sample. This explains partly the similarity in the morphology of the K-band spectra. This has the important consequence that all stars lie almost on a vertical line in the HR diagram (see Sect.\ \ref{s_hr}). Such a feature can be used to trace the age of the cluster. One can also note that the average temperature of the O4--6 supergiants is similar to the calibrated values of \citet{msh05} for O5--5.5I stars. 

\item[$\diamond$] \textit{The stars are luminous}: all the stars we analyzed have \lL $> 5.7$, some of them reaching \lL $= 6.35$. WN7--9h stars have, on average, \lL = 6.14, O4--6 supergiants have \lL = 5.84. Contrary to the case of effective temperatures, there is a clear luminosity difference between the classes of objects, WN7--9h stars being the most luminous. Luminosities in excess of $10^{6} L_{\odot}$ are indicative of very high mass stars. This characteristic will be further discussed in Sect.\ \ref{s_nature_wn7}. Early type supergiants/giants have luminosities in agreement with the expectations: the calibrations of \citet{msh05} indicate \lL = 5.78--5.95 for O4--6I stars.  

\item[$\diamond$] \textit{The stars are usually H rich}: all the stars have a ratio He/H $<$ 1.0. The WN7--9h stars are He enriched, but not excessively, with He/H between 0.1 and 1.0. Actually, only three WN7--9h stars have He/H $>$ 0.5: all the other WNLh stars have a lower He content. It is interesting to note that some WNLh stars do not seem to be He enriched (stars B1, F1, F9, F14, F16). All the other types of stars show a solar helium abundance (with the exception of star F10 which is slightly enriched -- He/H = 0.2). 

\item[$\diamond$] \textit{The stars show various degrees of chemical enrichment}: as expected, WNLh stars have the strongest nitrogen enrichment, with X(N) between 0.005 and 0.028. O supergiants, including the O4--6If$^{+}$ stars, have solar or slightly enriched N abundances (X(N) = 0.002--0.007). In parallel, WN stars are carbon deficient (X(C) $< 1.3 \times\ 10^{-3}$; exception: star F16) compared to O supergiants. This is a clear indication of CNO processing. 

\item[$\diamond$] \textit{Mass loss rates are stronger in WN stars}: all WN7--9h stars have \mdot $> 10^{-5}$ \myr (exception: star F16, \mdot $= 6.3 \times\ 10^{-6}$ \myr). The O4--6 supergiants have lower mass loss rates (in the range  $2-4.5 \times\ 10^{-6}$ \myr). 

\end{itemize}

From this general overview, one sees that the stars analyzed here share some common properties (\teff) but also are very different in terms of abundance patterns. This will be discussed in greater depth in Sect.\ \ref{s_chem}.

\begin{sidewaystable*}
\begin{center}
\caption{Derived stellar and wind parameters. The typical errors are: $\pm$3000 K on temperatures, $\pm$0.2 dex on \lL\ and $\log \dot{M}$, 100 \kms\ on terminal velocities and $\pm$30\%\ on abundances. Terminal velocities are adopted from \vinf\ = 2.6 $\times$ \vesc\ for O stars (except the two O4--6If$^{+}$ supergiants). \label{tab_res}}
\begin{tabular}{clrrrrrrrrrrrrrrrrr}
\hline\hline
Star   & ST         & T$_{*}$ & \teff & \lL  & R$_{*}$ & R$_{2/3}$ & M$_{K}$ & log \mdot & \vinf & He/H & X(C)  & X(N) & $\log Q_{H}$ & $\log Q_{\ion{He}{i}}$ & \mdot \vinf / (L/c)\\
       &            & [kK]     & [kK]   &   & [R$_{\odot}$] & [R$_{\odot}$] &  & [\myr]  & [\kms]& [\#] & [\%]  & [\%] &[s$^{-1}$]&[s$^{-1}$] & \\
\hline 		   				        								     
B1  & WN8--9h      & 32.2 & 31.7 & 5.95 & 30.5  & 31.5 & -6.00   & -5.00     & 1600  & 0.1 & $<$0.033 & 2.41  & 49.52 & 48.27 & 0.89\\
F1  & WN8--9h      & 33.7 & 33.2 & 6.30 & 41.6  & 43.0 & -6.75   & -4.70     & 1400  & 0.1 &    0.058 & 1.45  & 49.95 & 48.84 & 0.69\\
F2  & WN8--9h      & 34.5 & 33.5 & 6.00 & 28.1  & 29.9 & -6.07   & -4.72     & 1400  & 0.35& $<$0.015 & 1.43  & 49.67 & 48.57 & 1.32\\
F3  & WN8--9h      & 29.9 & 29.6 & 6.10 & 42.1  & 42.8 & -6.69   & -4.60     &  800  & 0.6 & $<$0.069 & 2.79  & 49.50 & 47.62 & 0.79\\
F4  & WN7--8h      & 37.3 & 36.8 & 6.30 & 33.9  & 34.8 & -6.81   & -4.35     & 1400  & 0.4 & $<$0.018 & 2.10  & 50.01 & 49.00 & 1.55\\
F5  & WN8--9h      & 35.8 & 32.1 & 5.95 & 24.6  & 30.6 & -6.34   & -4.64     &  900  & 0.8 & $<$0.011 & 1.95  & 49.67 & 48.55 & 1.14\\
F6  & WN8--9h      & 34.7 & 33.9 & 6.35 & 41.7  & 43.5 & -6.81   & -4.62     & 1400  & 0.2 &    0.046 & 1.14  & 50.04 & 49.00 & 0.74\\
F7  & WN8--9h      & 33.7 & 32.9 & 6.30 & 39.4  & 41.2 & -6.70   & -4.60     & 1300  & 0.3 &    0.011 & 1.86  & 49.91 & 48.77 & 0.81\\
F8  & WN8--9h      & 33.7 & 32.9 & 6.10 & 33.1  & 34.7 & -6.50   & -4.50     & 1000  & 1.0 & $<$0.023 & 1.64  & 49.74 & 48.55 & 1.24\\
F9  & WN8--9h      & 36.8 & 36.6 & 6.35 & 36.9  & 37.4 & -6.37   & -4.78     & 1800  & 0.1 &    0.042 & 1.46  & 50.06 & 49.10 & 0.66\\
F10 & O4--6If$^{+}$& 32.4 & 32.2 & 5.95 & 30.1  & 30.4 & -5.76   & -5.30     & 1600  & 0.1 &    0.170 & 0.39  & 49.41 & 47.99 & 0.44\\
F12 & WN7--8h      & 37.3 & 36.9 & 6.20 & 30.3  & 30.9 & -6.21   & -4.75     & 1500  & 0.2 & $<$0.013 & 2.26  & 49.90 & 48.92 & 0.83\\
F14 & WN8--9h      & 34.5 & 34.5 & 6.00 & 28.2  & 29.9 & -5.94   & -5.00     & 1400  & 0.1 &    0.130 & 0.49  & 49.68 & 48.64 & 0.69\\
F15 & O4--6If$^{+}$& 35.8 & 35.6 & 6.15 & 31.0  & 31.4 & -5.97   & -5.10     & 2400  & 0.1 &    0.067 & 0.49  & 49.80 & 48.81 & 0.67\\
F16 & WN8--9h      & 32.4 & 32.2 & 5.90 & 28.5  & 28.7 & -5.75   & -5.11     & 1400  & 0.1 &    0.416 & 1.46  & 49.37 & 47.95 & 0.68\\	    
F18 & O4--6I       & 37.3 & 36.9 & 6.05 & 25.5  & 26.1 & -5.58   & -5.35     & 2150  & 0.1 &    0.084 & 0.39  & 49.77 & 48.88 & 0.42\\
F20 & O4--6I       & 38.4 & 38.2 & 5.90 & 20.3  & 20.4 & -5.08   & -5.42     & 2850  & 0.1 &    0.169 & 0.30  & 49.59 & 48.73 & 0.57\\
F21 & O4--6I       & 35.8 & 35.5 & 5.95 & 24.7  & 25.1 & -5.44   & -5.49     & 2200  & 0.1 &    0.084 & 0.39  & 49.61 & 48.65 & 0.46\\
F22 & O4--6I       & 35.8 & 35.4 & 5.80 & 20.8  & 21.2 & -5.07   & -5.70     & 1900  & 0.1 &    0.127 & 0.39  & 49.46 & 48.48 & 0.30\\
F23 & O4--6I       & 35.8 & 35.4 & 5.80 & 20.8  & 21.2 & -5.08   & -5.65     & 1900  & 0.1 &    0.169 & 0.69  & 49.46 & 48.47 & 0.33\\
F26 & O4--6I       & 39.8 & 39.6 & 5.85 & 17.8  & 18.0 & -4.82   & -5.73     & 2600  & 0.1 &    0.127 & 0.40  & 49.58 & 48.76 & 0.34\\
F28 & O4--6I       & 39.8 & 39.6 & 5.95 & 19.9  & 20.1 & -5.06   & -5.70     & 2750  & 0.1 &    0.296 & 0.40  & 49.68 & 48.88 & 0.30\\
F29 & O4--6I       & 35.7 & 35.3 & 5.75 & 19.6  & 20.1 & -4.90   & -5.60     & 2900  & 0.1 &    0.253 & 0.30  & 49.34 & 48.32 & 0.64\\
F32 & O4--6I       & 40.8 & 40.5 & 5.85 & 16.9  & 17.2 & -4.73   & -5.90     & 2400  & 0.1 &    0.672 & 0.29  & 49.68 & 48.88 & 0.21\\
F33 & O4--6I       & 39.8 & 39.6 & 5.85 & 17.8  & 18.0 & -4.82   & -5.73     & 2600  & 0.1 &    0.127 & 0.39  & 49.58 & 48.76 & 0.34\\
F34 & O4--6I       & 38.1 & 37.4 & 5.75 & 17.3  & 17.9 & -4.78   & -5.77     & 1750  & 0.1 &    0.127 & 0.40  & 49.49 & 48.63 & 0.26\\
F35 & O4--6I       & 33.8 & 33.5 & 5.70 & 20.7  & 21.1 & -4.74   & -5.76     & 2150  & 0.1 &    0.296 & 0.20  & 49.26 & 48.11 & 0.37\\
F40 & O4--6I       & 39.8 & 39.5 & 5.75 & 15.8  & 16.1 & -4.58   & -5.75     & 2450  & 0.1 &    0.127 & 0.40  & 49.60 & 48.82 & 0.38\\
\hline
\end{tabular}
\end{center}
\end{sidewaystable*}

\subsection{Comparison to previous studies}
\label{s_comp_paco}

The only attempt to derive quantitative properties of the Arches
massive stars is by \citet{paco04}. The authors focused on five stars
(F3, F4, F8, F10 and F15) and used a similar technique to determine the
stellar and wind parameters. They relied on Keck/NIRSPEC spectra with
a high resolution ($\sim$ 23300) but a narrower spectral coverage than
our VLT/SINFONI data (only four windows centered on \heia, \heib,
\brg\ and \niiia\ were observed).

\begin{table*}
\begin{center}
\caption{Comparison between our results and the study of \citet{paco04} for the five stars in common. For each star, the first row gives our results, and the second one the results of \citet{paco04}. Only the main parameters are listed. \label{tab_comp_paco}}
\begin{tabular}{clrrrrrrrrrrrrrrrr}
\hline\hline
Star   & ST            & \teff & \lL  & $\log \frac{\dot{M}}{\sqrt{f}}$ & \vinf   & He/H & X(C)   & X(N)\\
       &               &  [kK] &      & [\myr]                          & [\kms]  &      & (\%)   & (\%)\\
\hline 	   
3      & WN8--9h       &  29.6 & 6.10 & -4.10                           &  800    & 0.6  & 0.069  & 2.8 \\
       &               &  27.9 & 6.01 & -4.17                           &  840    & 0.5  & 0.020  & 1.7 \\
4      & WN7--8h       &  36.8 & 6.30 & -3.85                           & 1400    & 0.4  & 0.018  & 2.1 \\
       &               &  33.2 & 6.22 & -4.07                           & 1400    & 0.57 & 0.030  & 1.4 \\
8      & WN8--9h       &  32.9 & 6.10 & -4.00                           & 1000    & 1.0  & 0.023  & 1.6\\
       &               &  30.9 & 6.27 & -3.80                           & 1100    & 0.67 & 0.020  & 1.6 \\
10     & O4--6If$^{+}$ &  32.2 & 5.95 & -4.80                           & 1600    & 0.1  & 0.170  & 0.4 \\
       &               &  30.7 & 6.27 & -4.87                           & $<$1000 & 0.33 & 0.080  & 0.6 \\
15     & O4--6If$^{+}$ &  35.6 & 6.15 & -4.60                           & 2400    & 0.1  & 0.067  & 0.5 \\
       &               &  29.5 & 5.77 & -4.54                           & $<$1000 & 0.33 & 0.150  & 0.6 \\
\hline
\end{tabular}
\end{center}
\end{table*}

The parameters derived in both the present study and the analysis of
\citet{paco04} are summarized in Table \ref{tab_comp_paco}. Generally,
there is a rather good agreement between both studies for the WN
stars. Note in particular the similar luminosities, mass loss rates,
terminal velocities and He and N abundances. We find effective
temperatures systematically larger (by $\sim$ 2000K), but the
difference is within the uncertainties (except for star F4). The
largest difference is found for the carbon abundance, although, with
one exception (star F3), the discrepancy is only a factor of 2 or
smaller. Here, we argue that \citet{paco04} did not cover the full
spectral range around 2.07\mum\ to observe the \civa\ lines (see their
Fig. 1), which is included in our SINFONI spectra. We have
consequently a larger number of diagnostics and we are able to better
derive the C content.

The differences are larger for the two O4--6If$^{+}$ stars. We think
the better quality of our spectra allows a better estimate of the
terminal velocity (the full width of \brg\ is well observed), the C
and N content (we unambiguously detect the \civa\ lines and the
\niiia\ doublet). The mass loss rates being similar, the wind
densities are not (due to the difference in \vinf), which partly
explains the different luminosities. The different effective
temperatures, well constrained by our well resolved \ion{He}{i} and
\ion{He}{ii} lines (especially \heiia) complete this explanation.

\section{HR diagram and cluster age}
\label{s_hr}

\begin{figure}
\includegraphics[width=9cm]{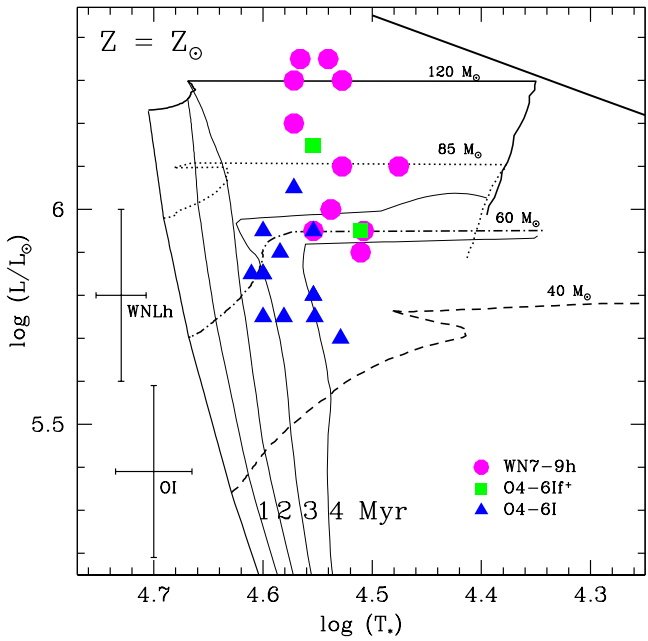}
\caption{HR diagram of the Arches cluster. Filled symbols are the stars analyzed in this work. The Geneva evolutionary tracks including rotation from \citet{mm05} are plotted, as well as isochrones. For clarity, only the first $\sim$ 4 Myr of the tracks are plotted. The solid line in the right upper part of the figure is the Humphreys-Davidson limit. The typical error on the position of the stars is shown at the bottom left. \label{hr_arches}}
\end{figure}

The stars analyzed in the present study are placed in the HR diagram
in Fig.\ \ref{hr_arches}. The Geneva evolutionary tracks including
rotation from \citet{mm05} are used to build the diagram. Isochrones
are also shown. As previously described, there is a clear difference
in the position of stars of different spectral types. The WN7--9h
stars (filled circles) are brighter than the normal O supergiants. The
extreme supergiants are intermediate. One can immediately conclude
that the WN7--9h stars of the Arches cluster are very massive stars:
only the 120 \msun\ evolutionary track is able to reach luminosities
larger than $10^{6} L_{\odot}$. Even the less luminous WN7--9h stars
are accounted for only by the tracks with $M > 60 M_{\odot}$. One
concludes that in the Arches cluster, the WN7--9h stars are the
descendent of stars more massive than 60 \msun. The position of the
extreme early supergiants (the O4-6If$^{+}$ stars) overlaps with the
position of the less luminous WN7--9h stars. It is thus likely that
they are closely related to them (see also the next sections).

From the position of the stars in the HR diagram, one can attempt to
estimate their age. For that, isochrones are indicated in Fig.\
\ref{hr_arches}. The most luminous WN7--9h stars are 2-3 Myr old. The
O supergiants (with the exception of the O4-6If$^{+}$ stars) seem to
span a slightly wider age range (2-4 Myr). There is however an overlap
between the brightest supergiants and the faintest WN7--9h stars. This
suggests that on average the most massive stars are slightly younger
than less massive stars (in the mass range 40-120 \msun): one can
clearly exclude an age of 3 Myr for the most luminous WN7--9h stars,
while some of the O supergiants with $L \sim 10^{5.8} L_{\odot}$ could
be $\sim$ 4 Myr old. This may be an indication that the most massive
stars formed at the end of the star forming event that gave birth to
the Arches cluster. This would be consistent with the scenario
according to which the most massive stars are the last to form in a
starburst event since their presence immediately imply a strong
negative feedback which removes material for star formation. We note
however that we are probing a small region of the HR diagram and this
would need to be confirmed by a deeper study of intermediate and low
mass stars. Within the uncertainty on the effective temperature and
luminosity of the O supergiants, one cannot exclude either that they
have the same age as the WNLh stars. Besides, binarity may change the
picture. If the most luminous stars were found to be binary stars, the
luminosity of each component would have to drop (by as much as 0.3 dex
in case of equally luminous companions). This would translate into an
older age for the stars, now closer to the 3Myr isochrones in Fig.\
\ref{hr_arches}. Hence, the suggestion of a late formation of the most
massive stars still needs to be confirmed, but is worth being
mentioned in view of the present results. We note also that in
NGC3603, a galactic cluster quite similar to the Arches, there exists
a population of pre-main sequence stars younger than the most massive
components \citep{stolte06,yohei07}. Clearly, a study of fainter
components of the cluster is required to confirm the suggestion of a
correlation between age and initial mass.

\citet{figer99} derived an age of 2$\pm$1 Myr for the Arches cluster
based on photometry of the massive components. A better estimate was
given by \citet{blum01} who used information on the spectral types in
combination to evolutionary models to constrain the age of the Arches
to the range 2-4.5 Myr. Finally, comparing the types of Wolf-Rayet
stars present in the cluster to the predictions of starburst models of
\citet{meynet95}, as well as using the detailed properties of one star
(F8), \citet{figer02} concluded that the age of the cluster is
2.5$\pm$0.5 Myr. Our determination, which relies on a more
quantitative basis and uses the most recent evolutionary tracks,
agrees nicely with this estimate (especially if only WR stars are
considered).

Finally, one should mention that the above results were obtained using
solar metallicity tracks. As we will see later, the metallicity in the
cluster might be slightly super-solar. In that case, the age we
derived would be only an upper limit. Indeed, for Z=2$\times$\zsun,
comparison to the corresponding evolutionary tracks show that
\textit{all} the stars we analyzed would be younger than 3 Myr.

\section{Nature of the most luminous stars}
\label{s_nature_wn7}

In this section we investigate the nature of the Arches stars studied here as well as the possible relation between the different types of stars (O, WNLh). We first focus on the chemical evolutionary status, then discuss the wind properties before drawing our conclusions.

\subsection{Chemical evolution}
\label{s_chem}

In the previous section we have seen that the stars classified WN7--9h
in our sample appear to be very massive stars. To better unravel the
nature of these objects, the analysis of their abundance pattern is a
powerful tool since it informs about their evolutionary status.

\begin{figure}
\includegraphics[width=9cm]{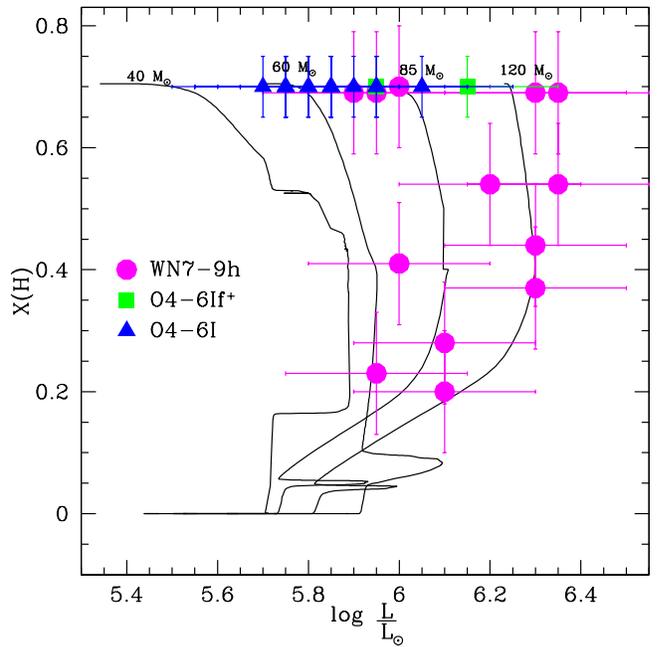}
\caption{Hydrogen mass fraction as a function of luminosity in the solar metallicity Geneva evolutionary models of \citet{mm05} (solid line) and as derived in the stars analyzed in this work (symbols). \label{xh_l}}
\end{figure}

In Fig.\ \ref{xh_l} we plot the hydrogen mass fraction as a function
of luminosity. The symbols have the same meaning as in Fig.\
\ref{hr_arches}. The relation from the Geneva evolutionary tracks are
overplotted (solid lines). One can immediately conclude that the
WN7--9h stars are the only ones of our sample showing H
depletion. However, even some of the WN7--9h stars appear not to be He
enriched (none of the O stars are H depleted). Overall, the H mass
fraction of WN7--9h stars ranges between 0.2 and 0.7. We interpret
this pattern as a sign that the Arches WNLh stars are objects which
left the main sequence recently, some of them being almost unevolved
in terms of H depletion. This is a very important conclusion, because
it means that some (and maybe all) of these stars are still core H
burning objects. This implies that these stars are young, consistent
with our age estimate (see Sect.\ \ref{s_hr}). The comparison with the
evolutionary tracks in Fig.\ \ref{xh_l} also indicates that they have
masses in the range 60--120 \msun, similar to what was inferred from
their position in the HR diagram.

\begin{figure}
\includegraphics[width=9cm]{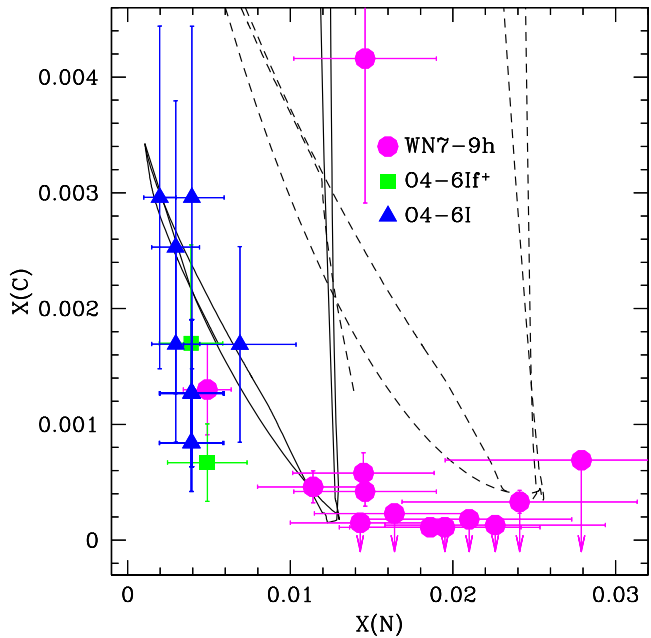}
\caption{Carbon mass fraction as a function of Nitrogen mass fraction in the Geneva evolutionary models of \citet{mm05} (solid line: solar metallicity; dashed line: twice solar metallicity) and as derived in the stars analyzed in this work (symbols). The evolutionary tracks plotted are for 20, 40 and 85 \msun. \label{xc_xn}}
\end{figure}

A quantitative confirmation of the young evolutionary status of the
Arches WNLh stars is given by Fig.\ \ref{xc_xn}. In this figure, the
carbon mass fraction (X(C)) is shown as a function of the nitrogen
mass fraction (X(N)). According to stellar evolution, X(C) decreases
in the earliest phases while X(N) increases when H is burnt through
the CNO cycle. Then when He burning starts, C is produced at the
expense of N (and He). Fig.\ \ref{xc_xn} reveals that the WN7--9h
stars are all carbon poor and N rich compared to the O stars of the
sample (with the exception of star 14 which is more similar to O
stars). In fact, they cluster in a rather small region of the
X(C)--X(N) diagram, which suggests that they show the pattern of CNO
equilibrium. Contrarily to Fig.\ \ref{xh_l}, there is a clear
distinction between WN7--9h and O stars which are more C rich and N
poor. This allows us to unambiguously state that the WN7--9h stars are
evolved objects but are still at the beginning of their post main
sequence evolution.

\begin{figure*}
 \centering
 \begin{minipage}[b]{8.8cm}
   \includegraphics[width=9cm]{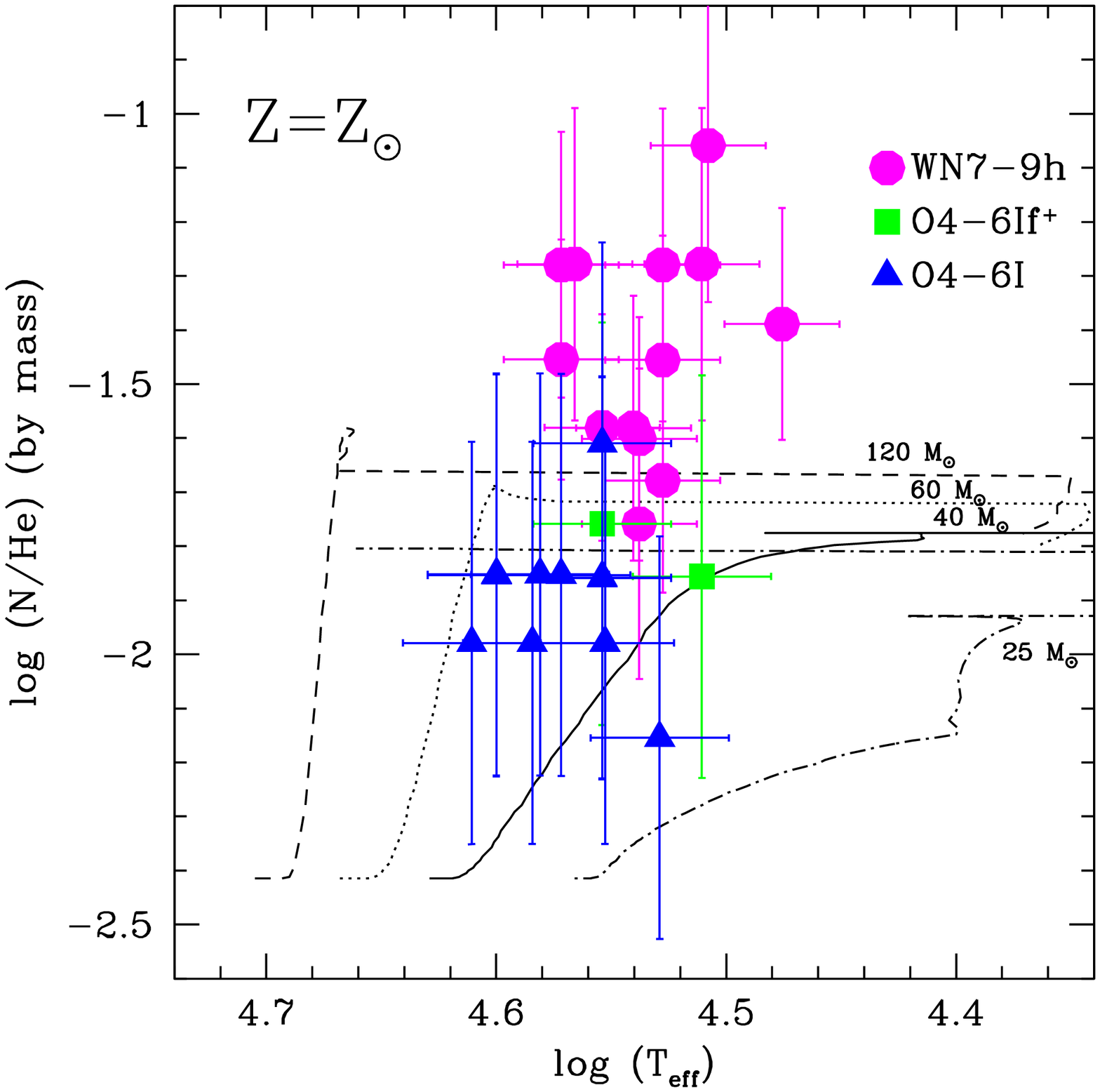}
 \end{minipage}
 \begin{minipage}[b]{8.8cm}
   \includegraphics[width=9cm]{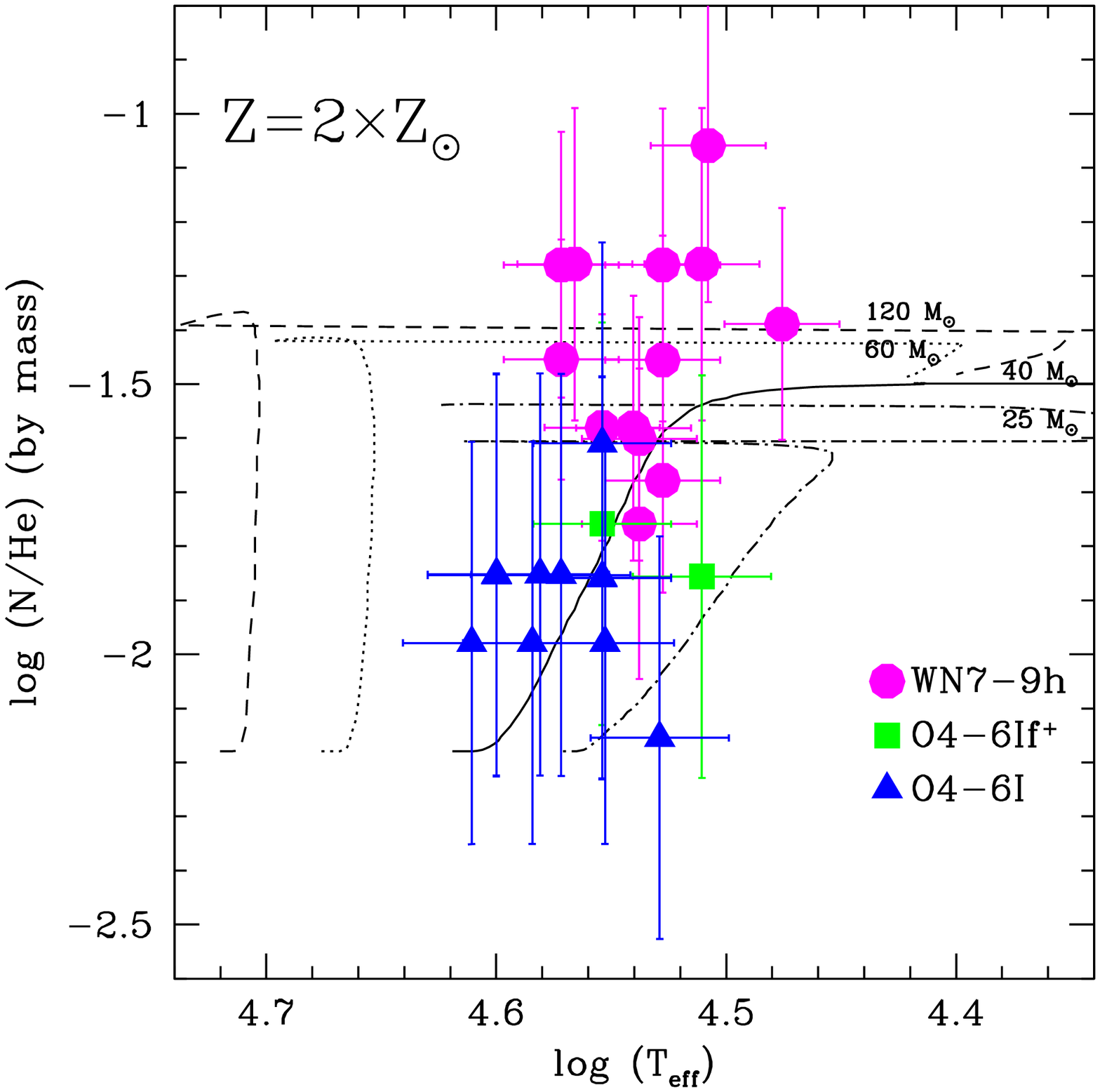}
 \end{minipage}
\caption{Logarithm of the ratio of nitrogen to helium mass fraction in the Geneva evolutionary models of \citet{mm05} and as derived in the stars analyzed in this work (symbols). \textit{Left}: solar metallicity evolutionary tracks. \textit{Right}: twice solar metallicity tracks.}
\label{NonHe_teff}
\end{figure*}

Fig.\ \ref{xc_xn} shows that although the trend of decreasing X(C)
with increasing X(N) predicted by the evolutionary models is
\textit{qualitatively} reproduced by the observations,
\textit{quantitatively} the agreement is not perfect. In particular,
the WN7--9h stars span a wider range in X(N) (0.005 to 0.028) than
expected from the models. In the models, at the level of X(C) seen in
the WN7--9h stars, X(N) should be around 0.013 according to the solar
metallicity track. The track at twice the solar metal content allows
X(N) as large as 0.025. But for the level of X(N) observed, the C
content in this track should be up to 4--5 times larger. An
explanation purely by a non solar metallicity is not
satisfying. Different initial rotational velocities cannot fully
explain this scatter in X(N) values. Between two stars with initial
rotational velocities of 0 and 300 km/s, the difference of maximum
value of the N mass fraction X(N) is 0.002. This is about one tenth of
the range spanned by the WNLh stars (from $\sim$ 0.01 to $\sim$0.03,
excluding star F14). Hence, a spread in initial rotational velocities
cannot be fully responsible for the observed scatter in X(N).

One can also speculate that the evolutionary models do not predict a
strong enough N enrichment in the early phases of evolution of massive
stars. One important ingredient which is still neglected in most
evolutionary models is magnetic fields. Recently, \citet{mm05b} have
shown that the presence of magnetic fields in massive stars could
favor solid body rotation and consequently chemical mixing. Their
Fig. 10 reveals that He and N abundances can be significantly
increased compared to non magnetic models. Interestingly, the effect
of magnetic fields on chemical enrichment are larger for older
stars. This is to be compared to the larger spread in X(N) for more
evolved stars (WNLh compared to O supergiants) in Fig.\
\ref{xc_xn}. We do not claim that magnetic field can explain all the
trends seen in this figure, but it might be an important
ingredient. In that context, it is worth noting that \citet{trundle05}
report a similar discrepancy between derived (by atmosphere modeling)
and predicted (by evolutionary tracks) N abundances for B supergiants
in the SMC.

To further investigate the chemical evolutionary status of the Arches
stars, we have plotted in Fig.\ \ref{NonHe_teff} the ratio of N to He
abundance (by mass) as a function of effective temperature. Here
again, the Geneva evolutionary tracks for solar and twice solar
metallicity are overplotted. Theoretically, the N/He ratio probes the
first phases of evolution. Indeed, while both N and He are produced
during the H burning phase, the relative increase of the abundances
relative to the initial values is larger for N than for He, simply
because He is already a main element in the star while N is not. In
practice, the N mass fraction changes by an order of magnitude, while
the He mass fraction increases by only a factor of 2--3 \citep[see for
example Fig.\ 16 of][]{mm03}. The N/He ration thus evolves from the
initial value to a value corresponding to the CNO equilibrium. Stars
which have not yet reached this equilibrium will show intermediate
values of N/He. Focusing on the left part of Fig.\ \ref{NonHe_teff}
corresponding to the case of solar metallicity tracks, we see again a
clear difference between the WNLh and the O stars: the latter appear
to lie on evolutionary tracks where the N/He ration is still rising,
while the former are all in the region where N/He is constant. This
confirms our previous finding that the WNLh stars are core H burning
stars which have reached the CNO equilibrium. On the contrary, the O
stars are still in the process of reaching this equilibrium. An
important discrepancy between evolutionary tracks and observed stars
is the very large values of N/He in the Arches WNLh stars compared to
the tracks: for some stars, no track seems to be able to reproduce
their position. Even tracks for stars more massive than 120 \msun\
(the most massive star for which evolutionary tracks exist) would not
help, since all the tracks between 60 and 120 \msun\ seem to predict
about the same amount of N/He at equilibrium. More massive stars would
show the same values.  If we turn to the right part of Fig.\
\ref{NonHe_teff} where the twice solar metallicity tracks are shown
together with the derived stellar properties of the Arches stars, the
situation dramatically improves. All stars can now be represented by
the theoretical tracks, even the WN7--9h stars with the largest N/He
ratios. The ones with the lowest ratios now appear not to have reached
completely the CNO equilibrium. The reason for the largest theoretical
N/He ratio at twice solar metallicity is that between $Z_{\odot}$ and
$Z=2\times Z_{\odot}$ the initial He fraction barely changes, while
the N content is much larger. The initial N/He ratio is thus
larger. Does that mean that the Arches cluster metallicity is super
solar? Although Fig.\ \ref{NonHe_teff} makes a good case for it, we
have seen that the interpretation was a bit different for Fig.\
\ref{xc_xn} for which a super solar metallicity did not fully explain
the observed trend. The question of the Arches cluster metallicity
will be debated in Sect.\ \ref{s_Z}.

\begin{figure}
\includegraphics[width=9cm]{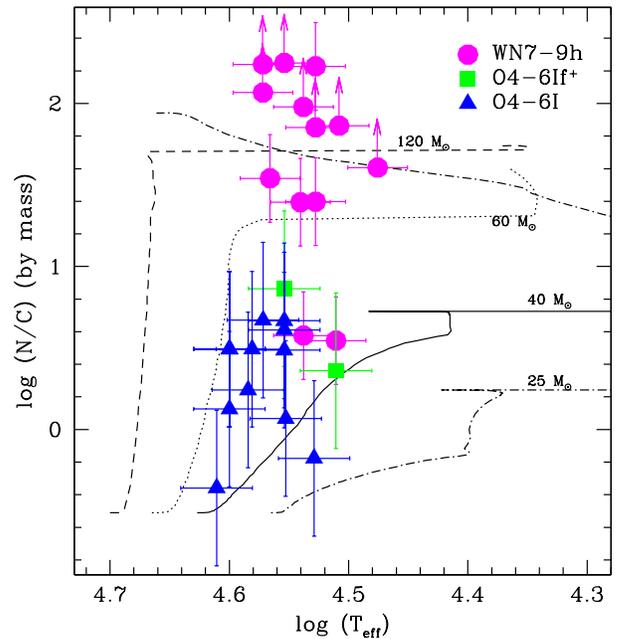}
\caption{Logarithm of the ratio of nitrogen to carbon mass fraction in the Geneva evolutionary models of \citet{mm05} and as derived in the stars analyzed in this work (symbols). \label{NonC_teff}}
\end{figure}

\begin{figure}
\includegraphics[width=9cm]{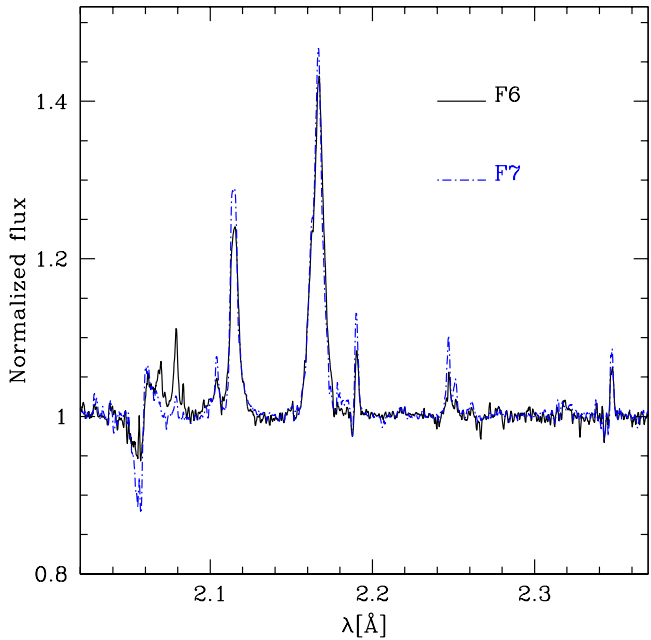}
\caption{Comparison between the spectra of stars F6 and F7 showing that stars with similar main lines and thus similar parameters (\teff, \lL, \mdot) can have different C and N abundances, as revealed by the different strengths of the \civa\ and \niiia\ lines. \label{comp_f67}}
\end{figure}

A final plot which combines the two previous ones is shown in Fig.\
\ref{NonC_teff}. It displays the abundance ratio N/C as a function of
\teff. Here again, this ratio reaches a maximum value when the stars
is at the CNO equilibrium. The WNLh stars and the O stars are clearly
separated, the former being more N rich--C poor than the latter. Note
however the existence of some overlap, two WNLh stars having N/C
ratios similar to O stars. We note that some WN7--9h stars still show
N/C ratios larger than predicted. Using evolutionary tracks with
$Z=2\times Z_{\odot}$ does not really improve the situation in this
case, since the initial N and C abundances are changed in a very
similar way when Z is increased, so that the initial N/C ratio does
not vary a lot.  Finally, it is interesting to note that the scatter
of the N/C ratio seen in Fig.\ \ref{NonC_teff} among WN7--9h stars is
real. An observational demonstration is made in Fig.\ \ref{comp_f67}
where we see that the two stars F6 and F7 have very similar spectra
(and thus similar parameters, see Table \ref{tab_res}) but different
\civa\ and \niiia\ lines. This means that stars at the same position
in the HR diagram can be in slightly different states of chemical
evolution.

From this analysis, we can safely conclude that the WN7--9h stars in
the Arches are core H burning objects showing products of the CNO
equilibrium at their surface. They are clearly distinct from the less
evolved O supergiants.

\subsection{Winds}
\label{s_winds}

In Fig.\ \ref{mdot_l} we show the mass loss rates of the Arches stars
studied here as a function of luminosity (left). It is clear that the WNLh
stars have stronger winds than the O stars. Indeed, although mass loss
increases with luminosity as predicted by the theory of radiation
driven winds \citep{cak75}, there is a separation between the O and
WNLh stars for the luminosity range 5.7--6.2 in which both types of
stars are found. This is another indication that WN7--9h stars are
more evolved than O stars. It is also very interesting to note that
the extreme supergiants (O4--6If$^{+}$ stars) seem to have mass loss
rates intermediate between normal O supergiants and WN7--9h stars. As
we will see in Sect.\ \ref{s_wrapup}, this is an indication of an
evolutionary link between early O supergiants and WNLh stars.

\begin{figure*}
 \centering
 \begin{minipage}[b]{8.8cm}
   \includegraphics[width=9cm]{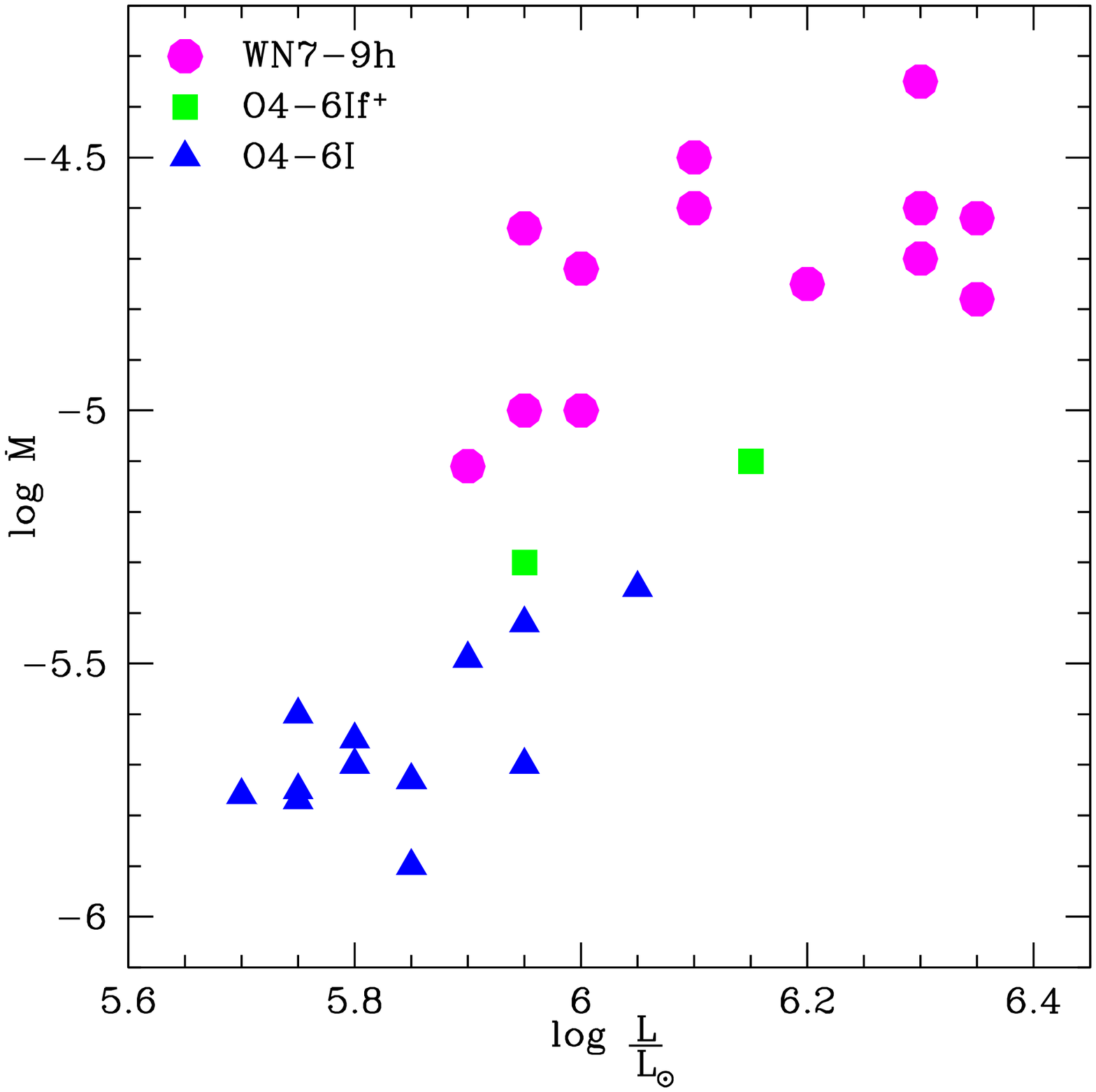}
 \end{minipage}
 \begin{minipage}[b]{8.8cm}
   \includegraphics[width=9cm]{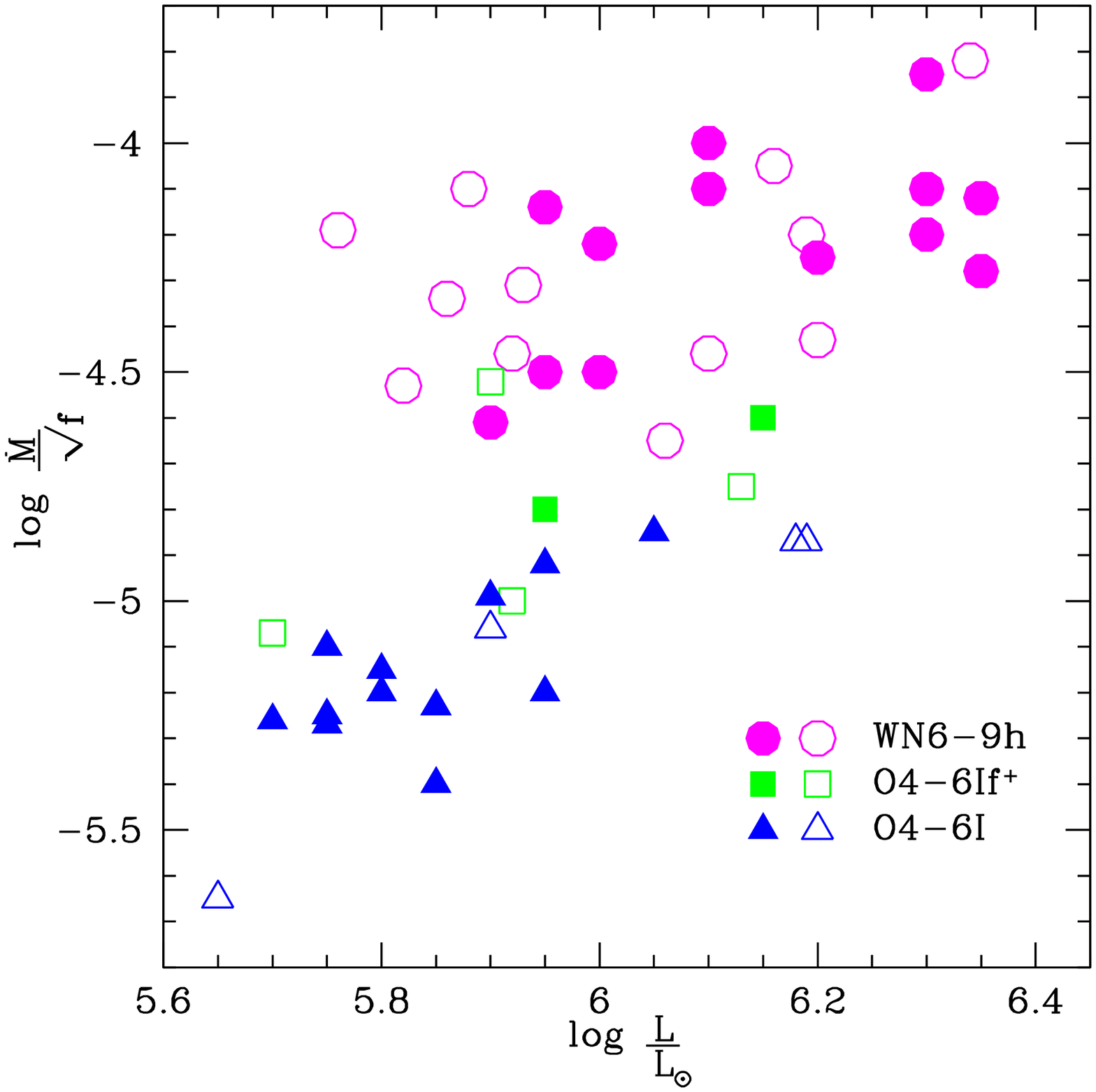}
 \end{minipage}
\caption{\textit{Left}: Mass loss rates as a function of luminosity for the Arches cluster stars. \textit{Right}: clumping corrected mass loss rates for the Arches cluster stars (filled symbols) as well as comparison objects (open symbols): early type (extreme) supergiants from \citet{herrero02,repolust05,jc05} and WN6-9h stars from \citet{cb97,cd98}}
\label{NonHe_teff}
\end{figure*}

\begin{figure}
\includegraphics[width=9cm]{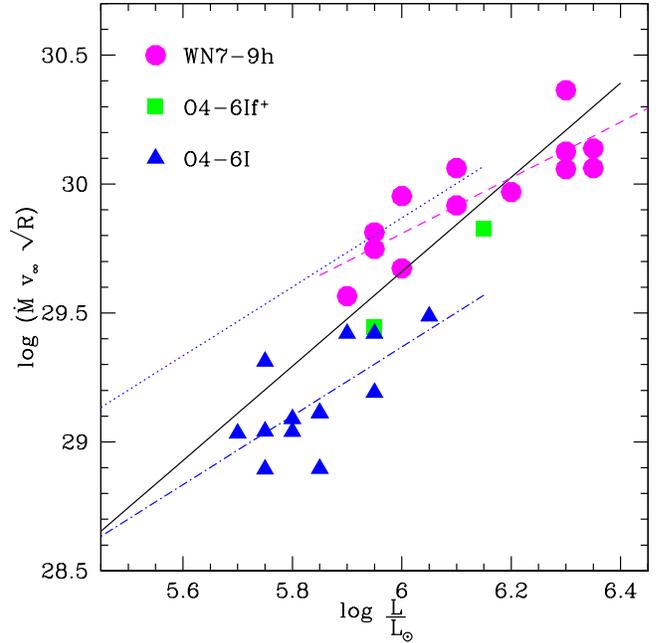}
\caption{Modified wind momentum -- luminosity for the Arches cluster stars. The solid line is the theoretical prediction from \citet{vink00}. The red dashed line is the linear fit to the O stars, excluding the O4--6If$^{+}$ supergiants. The blue dot dashed line is the fit to the WN7--9h stars. The dotted line is our WLR for O stars if we use unclumped mass loss rates. See text for discussion. \label{wlr_arches}}
\end{figure}

On the right part of Fig.\ \ref{mdot_l}, we show the clumping
corrected mass loss rates ($\dot{M}/\sqrt{f}$) as a function of
luminosity for our sample stars as well as for comparison objects. We
see that \textit{qualitatively}, the mass loss rates we derive are
consistent with previous studies of Galactic stars. Given the scatter
in \mdot\ among the various types of stars, it is not possible to make
a quantitative comparison. Instead, the so-called modified wind
momentum -- luminosity relation, or WLR, is a better tool. The
modified wind momentum, $\dot{M} v_{\infty} \sqrt{R}$ ($R$ being the
stellar radius), is expected on theoretical grounds to depend only on
luminosity \citep[e.g.][]{kp00}. Fig.\ \ref{wlr_arches} shows the
relation for the Arches cluster stars. On average, all stars follow
the qualitative increase of the modified wind momentum as a function
of luminosity. It is very important to note that this is also true to
a large extent for the WN7--9h stars: they clearly show a correlation
between modified wind momenta and luminosity. More specifically, we
have:

\begin{equation}
\log \dot{M} v_{\infty} \sqrt{R} = \left\{ \begin{array}{l}
 21.35(\pm 2.54)+1.34(\pm 0.44)  \log \frac{L}{L_{\odot}} \textrm{\hspace{0.2cm} OI} \\
 23.33(\pm 1.28)+1.08(\pm 0.21) \log \frac{L}{L_{\odot}} \textrm{\hspace{0.2cm} WNLh}
 \end{array} \right.
\end{equation}

The existence of a WLR for WN7--9h stars is a strong indication that
radiative acceleration plays an important in driving their winds. This
is a fundamental result, since the question of whether or not
radiative acceleration is efficient enough to produce the large WR
stars outflows is not entirely settled yet (see Crowther at al.\ 2007
for a recent review). Recent theoretical simulations by
\citet{goetz05} and \citet{vink05} convincingly indicate that
radiative acceleration might explain the large mass loss rates of WR
stars. Here, we provide an observational evidence that at least
qualitatively, the O and WNLh stars winds in the Arches cluster rely
on the same physics.  Another argument in favor of line driving for
the winds of the WN7--9h stars comes from the values of $\eta =
\dot{M} \times\ v_{\infty} /(L/c)$ listed in the last column of Table
\ref{tab_res}. We see that for these stars, it is close to 1, meaning
that radiation alone is in principle able to drive the wind, even in
the single scattering limit.

The WLR followed by the WN7--9h stars is systematically shifted
towards higher values (by $\sim$ 0.4 dex) compared to the O stars
relation. This is not surprising since such an effect is also seen
among O stars with different luminosity classes as well as between O,
B and A stars \citep{repolust04,kud99}. This is usually interpreted as
a change in the number of lines effectively driving the acceleration
\citep[see discussion in][]{kp00}. We also note that the WLR we derive
is flatter for WN7--9h stars: the slope is 1.34 for O stars, and 1.08
for WN7--9h stars. Given the errors, we cannot exclude however that
the slopes are similar. Note that if we calculate the slope of the WLR
regardless of the spectral types, i.e. including all stars of our
sample, we get a value of 2.00$\pm$0.19.

In Fig.\ \ref{wlr_arches}, we have also plotted the theoretical
relation of \citet{vink00} for O stars. Its slope is 1.83$\pm$0.04. We
see that given our limited sample and the errors, our WLR slope for O
stars is rather similar. The absolute position of the WLR is however
different. This is a known effect usually attributed to the use of
clumping in our models, while the models of \citet{vink00} are
homogeneous. Note however that hydrodynamical confirmations of this
effect are still lacking. Including clumping in atmosphere models
leads to systematically lower values of $\dot{M}$ compared to studies
with homogeneous winds, the difference being $\sqrt{f}$ with $f$ the
clumping factor. In our case, we use $f=0.1$ so that we can expect a
shift of 0.5 dex in the WLR of O stars (see dotted line in Fig.\
\ref{wlr_arches}). We see that if we correct our derived relation by
this amount, we end up slightly above the relation of \citet{vink00},
by a factor $\sim$ 0.2 dex (the difference between their theoretical
relation and ours is $\sim$ 0.2--0.3 dex depending on the
luminosity). How can we explain this remaining discrepancy?  One might
argue that our value of the clumping factor, assumed to be 0.1, is not
appropriate for O stars. Our relation would be consistent with that of
\citet{vink00} if we had chosen $f \sim 0.2-0.3$, a value that we
cannot discard from our modeling. As previously recalled, we note
however that recent studies indicate smaller $f$ for O supergiants
\citep{paul02,hil03,jc05}. Another interesting possibility to explain
the difference between our and Vink et al.'s WLR is that we see the
effects of high metal content. Mass loss rates of O stars are expected
to scale as $Z^{0.85}$ \citep{vink01}. If the difference we observe
between the theoretical relation of \citet{vink00} and our derived
relation for O stars was due to such an effect, this would mean that
the metallicity of the Arches stars should be $Z/Z_{\odot} =
10^{\frac{0.2}{0.85}} \sim 1.7$. This is an intriguing possibility
that we will discuss further in Sect.\ \ref{s_Z}. Finally, we should
mention that the prediction of \citet{vink00} might not be
correct. However, several recent studies seem to confirm its validity,
at least at high luminosities \citep{markova04,mokiem06}.

To conclude this section on the wind properties of the Arches stars,
we summarize in Table \ref{tab_comp_lang} the mass loss rates of the
stars analyzed in the present study which have been observed at radio
wavelengths by \citet{lang05}. From these detections, the authors
derive values of the mass loss rates using the standard relation of
\citet{wb75}. Since this relation implies that stellar winds are
homogeneous, we have to compare the mass loss rates of \citet{lang05}
to our clumping corrected $\dot{M}$
(i.e. $\frac{\dot{M}}{\sqrt{f}}$). \citet{lang05} also use a different
distance (8.0kpc instead of 7.62kpc) and assume $\mu = 2.0$ to derive
their mass loss rates. We also have to correct for this. The initial
and corrected values of \mdot\ of \citet{lang05} are given in columns
3 and 4 of Table \ref{tab_comp_lang}. For stars 3, 5, 8 and 18, we see
that both our and the radio determination are consistent within a
factor of 2. For the remaining stars (1, 2, 4, 6), the differences can
be as large as almost an order of magnitude. Star F6 is suspected to
be the counterpart of the X-ray source A1N \citep{ly04}, indicating
that it might be a non-thermal emitter (colliding winds). Besides,
\citet{lang05} report that star F6 is variable. Hence, differences
between our determination and the radio mass loss rate is not
surprising. For the three remaining stars (F1, F2 and F4), we have no
explanation so far of the disagreement. One might invoke binarity
(although our spectra exclude the presence of a spectroscopic
companion) or crowding which may affect the radio determination (the
resolution -- $0.42\arcsec \times 0.17\arcsec$ -- being lower than in
our SINFONI data -- 0.20\arcsec).

Excluding star F6 (suspected binary), it is interesting to note that
the radio mass loss rates are systematically lower than the IR
unclumped ones (exception: F18). This effect was noted by
\citet{figer02} for one star (F8). In our case, he difference is 0.36
dex, with a rather large dispersion (0.32). Since the ratio of
unclumped mass loss rates is virtually similar to the ratio of wind
densities (recall that the $\dot{M}$ diagnostics are in fact density
indicators), this difference might tell us something about the
different clumping factors in the IR and radio emitting
regions. Density is $\propto \dot{M} / \sqrt{f}$, where $\dot{M}$ is
the ``true'' mass loss rate. Hence, the ratio of radio to IR unclumped
mass loss rates is $\propto \sqrt{f_{IR}/f_{radio}}$ where $f_{IR}$
($f_{radio}$) is the clumping factor in the IR (radio) emitting
region. With the observed trend, it seems that clumping is stronger in
the IR than in the radio emitting region (by a factor of $\sim$
5). This is in agreement (qualitatively and quantitatively) with the
recent findings of \citet{puls06} and with the theoretical predictions
of \citet{runacres02}. This might partly explain the systematic
differences seen in Table \ref{tab_comp_lang}.

\begin{table}
\begin{center}
\caption{Clumping corrected mass loss rates for the stars in common between the present sample and the sample of \citet{lang05}.   \label{tab_comp_lang}}
\begin{tabular}{crrr}
\hline\hline
Star       & $\log \frac{\dot{M}}{\sqrt{f}}$ & $\log \dot{M}$ (Lang et al.) & $\log \dot{M}$ (Lang et al.) corrected\\
\hline 	   
1          & -4.20                           & -4.72                      & -4.80                               \\
2          & -4.22                           & -4.72                      & -4.66                               \\
3          & -4.10                           & -4.37                      & -4.47                               \\
4          & -3.85                           & -4.72                      & -4.64                               \\
5          & -4.14                           & -4.43                      & -4.44                               \\
6          & -4.12                           & -3.65                      & -3.66                               \\
8          & -4.00                           & -4.34                      & -4.27                               \\
18         & -4.85                           & -4.72                      & -4.62                               \\
\hline
\end{tabular}
\end{center}
\end{table}

\subsection{Nature and evolution of the most luminous stars}
\label{s_wrapup}

The discussion in the last sections has lead to a clear picture for
the nature of the WN7--9h stars in the Arches cluster: \textit{they
are very massive ($60<M<120 \msun$) post main sequence objects still
in the H burning phase and have reached the CNO equilibrium}. They
clearly separate from the rest of the O stars studied here. As such,
they are very reminiscent of the H rich WN stars in the core of the
NGC 3603 and R136 clusters \citep{cd98}.

From the analysis of several Galactic WN stars, \citet{paul95} built
the following evolutionary sequence for stars more massive than 60
\msun:

O $\rightarrow$ Of $\rightarrow$ WNL+abs $\rightarrow$ WN7 ($\rightarrow$ WNE) $\rightarrow$ WC $\rightarrow$ SN

\noindent where WNL (WNE) stands for WN late (early), i.e. WN6-9
(WN2-5) stars. \citet{langer94} describe a similar sequence except
that 1) they include a Luminous Blue Variable phase after the H-rich
(=WNL+abs) phase, and 2) they do not explicitely list WN7 types in
their scenario but replace them by the term H-poor WN. However, they
tentatively identify H-rich WN stars as core H burning objects. They
also note that these objects are the most luminous WN stars.

The Arches WNLh stars nicely fit the global picture drawn in these
scenarios: they are very luminous, young, H rich objects clearly still
in the H burning phase and with masses in excess of 60 \msun. Besides,
our detailed analysis of the C and N abundance patterns of these stars
quantitatively strengthens the conclusion that they are in a
relatively early evolutionary state.  The direct link between O and
WNLh stars is also confirmed by our analysis. The O supergiants of our
sample all appear to be less evolved than the WNLh stars (see Fig.\
\ref{xh_l}, \ref{xc_xn}, \ref{NonHe_teff} and \ref{NonC_teff}). And
the two O4-6If$^{+}$ extreme supergiants have properties intermediate
between normal supergiants and WNLh stars: this is best seen in Fig.\
\ref{NonHe_teff} where they bridge the two latter classes of
objects. Note however that we cannot state that there is a direct link
between specific sub-classes of stars in the Arches cluster, namely
between O4-6 supergiants and WNLh stars: Fig.\ \ref{hr_arches} reveals
that the former probably have lower initial masses than the
latter. But we can safely conclude that WNLh stars have early O stars
as progenitors, since they should evolve from more massive (and
consequently hotter) O stars than the O supergiants of our sample. In
practice, the Arches WNLh stars should be the descendant of O2-4
stars. We note that \citet{cb97} argued for a direct link between O8If
and WN9ha stars. However, this was based on the study of only 3 stars,
compared to 28 in the present paper. The WN9ha of their study had in
addition a lower luminosity and temperature than the Arches WN7--9h
stars, and might thus be a different, initially less massive type of
WN9h star. Consequently, we do not think that our results are in
contradiction with \citet{cb97}, but rather that they refer to
different kinds of stars.

In conclusion, our findings strongly support the scenario according to
which, in the Arches cluster, the most massive O stars evolve into
extreme supergiants and then into H-rich WNL stars.

\section{Stellar metallicity}
\label{s_Z}

\begin{figure}
\includegraphics[width=9cm]{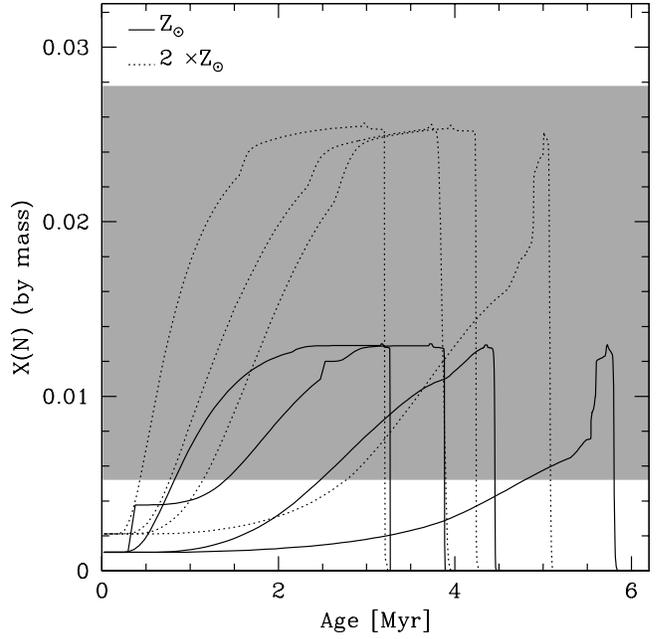}
\caption{N mass fraction as a function of age in evolutionary models for solar (solid line) and twice solar (dotted line) metallicity. The Geneva evolutionary tracks including rotation of \citet{mm05} are used. Tracks for M=120, 85, 60 and 40 \msun\ are shown from from left to right. The shaded area indicates the range of X(N) covered by the WN7-9h stars analyzed in the present study. \label{n_time}}
\end{figure}

\begin{figure}
\includegraphics[width=9cm]{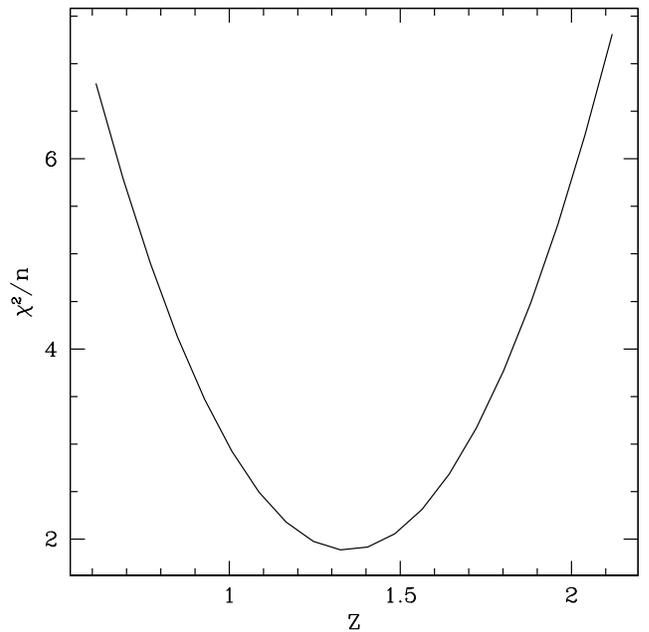}
\caption{Determination of metallicity from the N mass fraction of WNLh stars. $\chi^{2}$ is defined in Eq.\ \ref{def_chi2}. $\chi^{2}/n$ is shown as a function of metallicity. $n$ is the number of free parameters (12 is the present case). If the scatter in X(N) is purely statistical, then a slightly super-solar metallicity (Z=1.3--1.4\zsun) is preferred. \label{chi2_N_Z}}
\end{figure}

In the previous sections we have seen that several elements pointed
towards a super solar metallicity for the Arches cluster
stars: the N enrichment might be too large to be accounted for by
solar metallicity evolutionary models, and the winds might be stronger
than expected for a solar composition. This is somewhat in
contradiction with the recent results of \citet{paco04} who favored a
solar metallicity for the five stars they analyzed. Their
determination was based on the interesting finding that the N mass
fraction reaches a maximum in evolutionary tracks when the star is in
the WN phase. This maximum does not depend on the initial mass, but is
sensitive to the initial global metallicity. In practice, comparing
the mass fraction of a sample of WN stars to such tracks should then
constrain the metallicity. \citet{paco04} used the three WN stars they
analyzed to make such an estimate. They found that a solar metallicity
was preferred. In Table.\ \ref{tab_comp_paco} we have shown that for
these three stars, we find similar X(N) for one (F8), and slightly
larger values than \citet{paco04} for the other two (F3 and F4). Fig.\
\ref{n_time} is the figure Najarro et al.\ used to estimate Z in the
Arches cluster, but now using the 13 WN stars of our sample: the
shaded area corresponds to the range of X(N) covered by these 13
stars. From this, we see that a wide range of metallicity is
possible. If real, this can be attributed to two factors. First,
there may be a scatter in the initial metallicity of the Arches
stars. However, a difference of a factor of 2 seems quite large. The
second possibility is that the WN stars might not all have reached the
phase of their evolution were X(N) is maximum. In that sense, Fig.\
\ref{n_time} provides only a lower limit on the metallicity. Note that
this effect most likely influences the results when a small number of
stars is used. One might also wonder whether the scatter we
see is not purely statistical. If we assume it is the case, we can run
a $\chi^{2}$ analysis to find the preferred metallicity, using

\begin{equation}
\chi^{2}(Z) = \sum \frac{(X_{i}-X_{max}(Z))^{2}}{\sigma_{i}^{2}}
\label{def_chi2}
\end{equation}

\noindent where $X_{i}$ are the individual N mass fraction,
 $\sigma_{i}$ the associated uncertainties, and $X_{max}(Z)$ the
 maximum N mass fraction reached at a given metallicity.  The
 evolution of $\chi^{2}/n$ with $Z$ is shown in Fig.\ \ref{chi2_N_Z}
 ($n$ being the number of free parameters, equal to 12 in our case --
 13 stars included and one fitted parameter). We see that there is a
 clear minimum for Z=1.3--1.4 \zsun, and that a twice solar
 metallicity is clearly excluded. The minimum $\chi^{2}/n$ is 1.88, so
 a unique metallicity is not completely satisfactory to explain the
 distribution of $X(N)_{max}$. One has to keep in mind that this
 analysis is also correct only if the evolutionary models correctly
 predict the evolution of surface abundances.

Altogether, our results tend to favor a slightly super solar
metallicity. \citet{paco04} concluded that Z=\zsun\ was preferred. The
difference is likely due to the use of different evolutionary tracks:
we use the recent tracks from \citet{mm05}, while Najarro et al.\ use
the obsolete tracks from \citet{schaller92}. Since for a given
metallicity, these old tracks have a larger maximum X(N), a lower
metallicity is derived for a given range of observed X(N)
\footnote{Note that the maximum value of X(N) is independent of the
rotation rate of the star.}. In our sample, the average value of X(N) in
WNLh stars is 1.70($\pm 0.60) \%$. The three stars of Najarro et al.\
have on average X(N)=$1.57(\pm 0.15) \%$. Hence, within the errors,
the values are similar. The different derived metallicity we find is
thus likely due to the use of different evolutionary tracks.

An important comment to make is that the approach described above
is only valid if the CNO abundances and the global metallicity Z scale
similarly, or stated differently if all the metals have the same
relative overabundance compared to the solar composition. Let us
assume we have a model in which the initial CNO content is larger than
solar, while all the other metals have solar abundances. The global
metallicity will then be slightly above solar. Let us consider a
second model in which the global metallicity is the same as in the
first one, but now all the elements have the same abundance excess
relative to the solar composition. In the two models, the global
metallicity is the same, but the first one will produce a much larger
N mass fraction. If we use this mass fraction to asses the global
metallicity using the method presented above, then we will
overestimate the global metallicity of the star.

Studies of abundances of individual elements not affected by stellar
evolution would help to better constrain the metallicity of the Arches
cluster stars. Unfortunately, very few lines are available for such a
purpose. The only one which could be of help is the \ion{Si}{iv} line
at 2.428 \mum. Interestingly, it turns out that to correctly reproduce
it in all the stars we have studied so far, we need a silicon
abundance between solar and twice solar. Given that the line is
located at the end of the K band where the S/N degrades rapidly and
where the spectrum normalization is less straightforward than at
shorter wavelengths, we claim that this is only an indication that
individual abundances of light metals might be super-solar. It is
nonetheless interesting that this would be consistent with the various
indications gathered so far. If we assume that the Fe content is about
solar as studies of red supergiants in the central cluster show
\citep{carr00,ramirez00}, it might well mean that there is an excess
of light metals relative to heavier metals in the Arches cluster. Such
a conclusion would thus weaken the results of the Z determination by
the method presented by \citet{paco04}. But this would be a very
attractive possibility, since this could indicate a larger $\alpha$/Fe
abundance ratio, which in turn is usually interpreted as the imprint
of a top-heavy IMF. A recent study of stellar abundances in cool
luminous stars by \citet{cunha07} concluded that O and Ca (two
$\alpha$ elements) were overabundant compared to Fe in the central
cluster. Our suggestion of a super-solar Si abundance in the Arches
cluster is consistent with their findings.

In conclusion, we tentatively suggest that in the Arches cluster, the
lightest elements most likely have a super-solar abundance while the
iron peak elements have a solar metallicity.

\section{Conclusions}
\label{s_conc}

We have presented a quantitative study of the most massive stars in
the Arches cluster. K--band spectra have been obtained with SINFONI on
the VLT. A detailed spectral classification has revealed the presence
of WN7--9h stars as well as O supergiants, including two extreme
OIf$^{+}$ stars. We have quantified the main stellar and wind
parameters of 28 stars using the atmosphere code CMFGEN. The main
results of our study are:

\begin{itemize}
   \item the massive star population of the Arches cluster is 2--4 Myr old. Although marginal, there seems to be the trend that the most massive stars are also the youngest: WNLh stars are 2--3 Myr old, while less massive O supergiants might have an age of up to 4 Myr. Initial masses as large as 120 \msun\ are derived for the WNLh stars from the HR diagram. 
   \item The WN7--9h stars are identified as core H burning stars which show chemical enrichment typical of the CNO equilibrium: they still contain a significant amount of hydrogen and show both N enhancement and C depletion. They have supergiants of spectral types earlier than O4-6 as progenitors.
   \item The level of N enrichment suggests either a super-solar initial content and/or a too low efficiency of N enrichment in the evolutionary models. The indication that Si might be overabundant by a factor 2 compared to the solar abundance argues in favor of a super-solar metallicity at least for the lightest metals.
   \item The properties of the Arches massive stars argue in favor of the evolutionary scenario of \citet{paul95} for the most massive stars: O $\rightarrow$ Of $\rightarrow$ WNL+abs $\rightarrow$ WN7
   \item The winds of the WN7--9h stars follow a well defined modified wind momentum -- luminosity relation. This is a strong indication that they are radiatively driven. It also seems that the winds are less clumped in the radio continuum emitting region than in the near-IR line emitting region, in agreement with \citet{puls06}.
\end{itemize}

In order to test the indication that the most massive stars are the
last to form, intermediate mass stars must be analyzed. The question
of the metallicity in the Galactic Center, and the Arches cluster in
particular, is clearly not answered yet. New observations/analysis of
cooler stars showing a larger number of metallic lines are needed. We
will present such a study in a forthcoming paper.

\begin{acknowledgements}
We thank F. Najarro for useful comments and for sharing his
\ion{O}{iii} model atom. This paper benefited from interesting
discussions with G. Meynet. FM acknowledges partial support from the
Alexander von Humboldt foundation. Finally, we thank an anonymous
referee for a careful reading of the manuscript and valuable comments.
\end{acknowledgements}

\bibliography{biblio.bib}

\begin{appendix}
\section{Best fits}
In this Section, we gather the figures showing the comparison of our best fit models with the observed spectra of our program stars.

\begin{figure*}
\includegraphics[width=19cm,height=25cm]{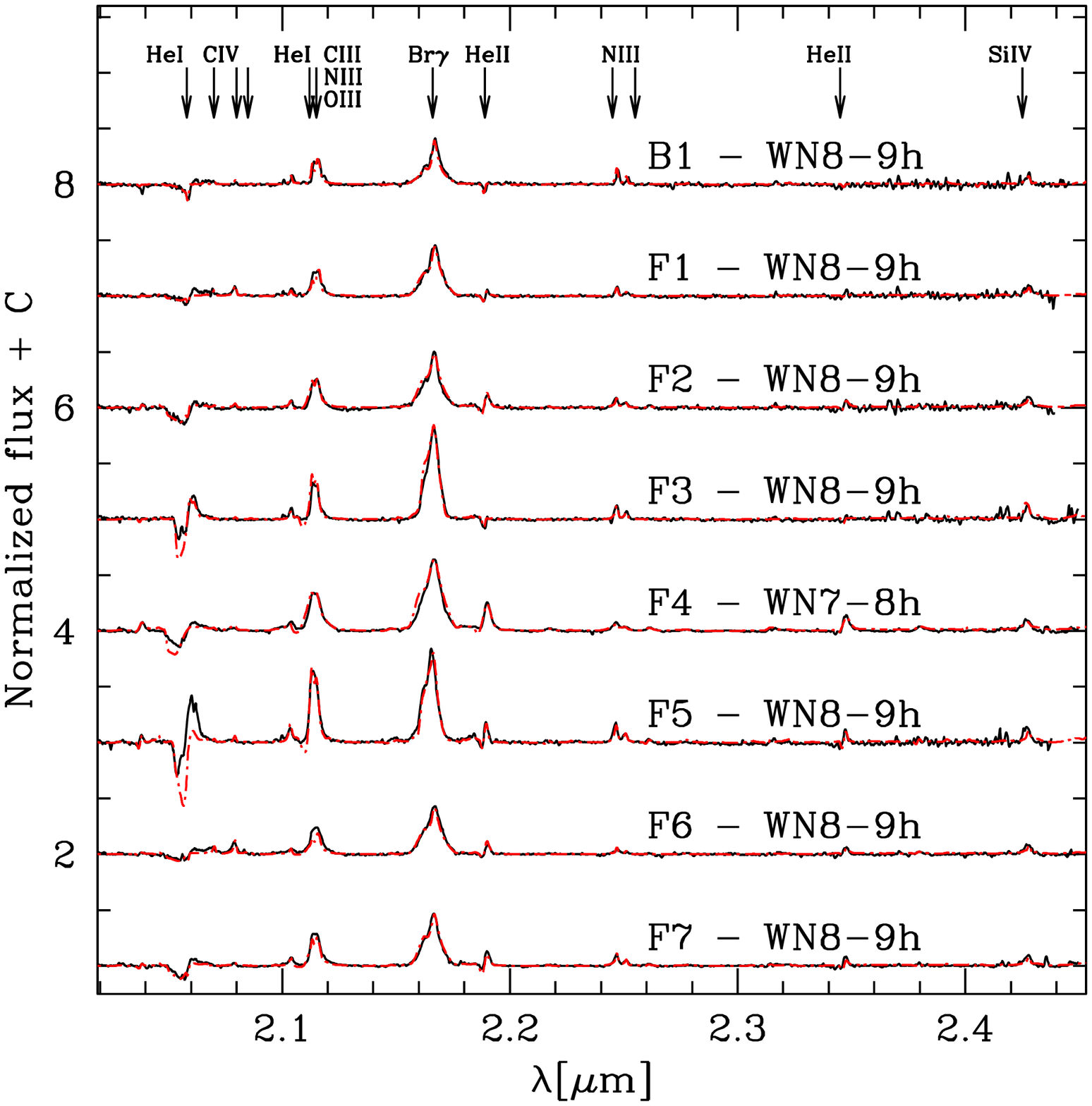}
\caption{Best fits (red dot-dashed lines) of the observed K-band spectra (black solid line). \label{fit_all_1}}
\end{figure*}

\begin{figure*}
\includegraphics[width=19cm,height=25cm]{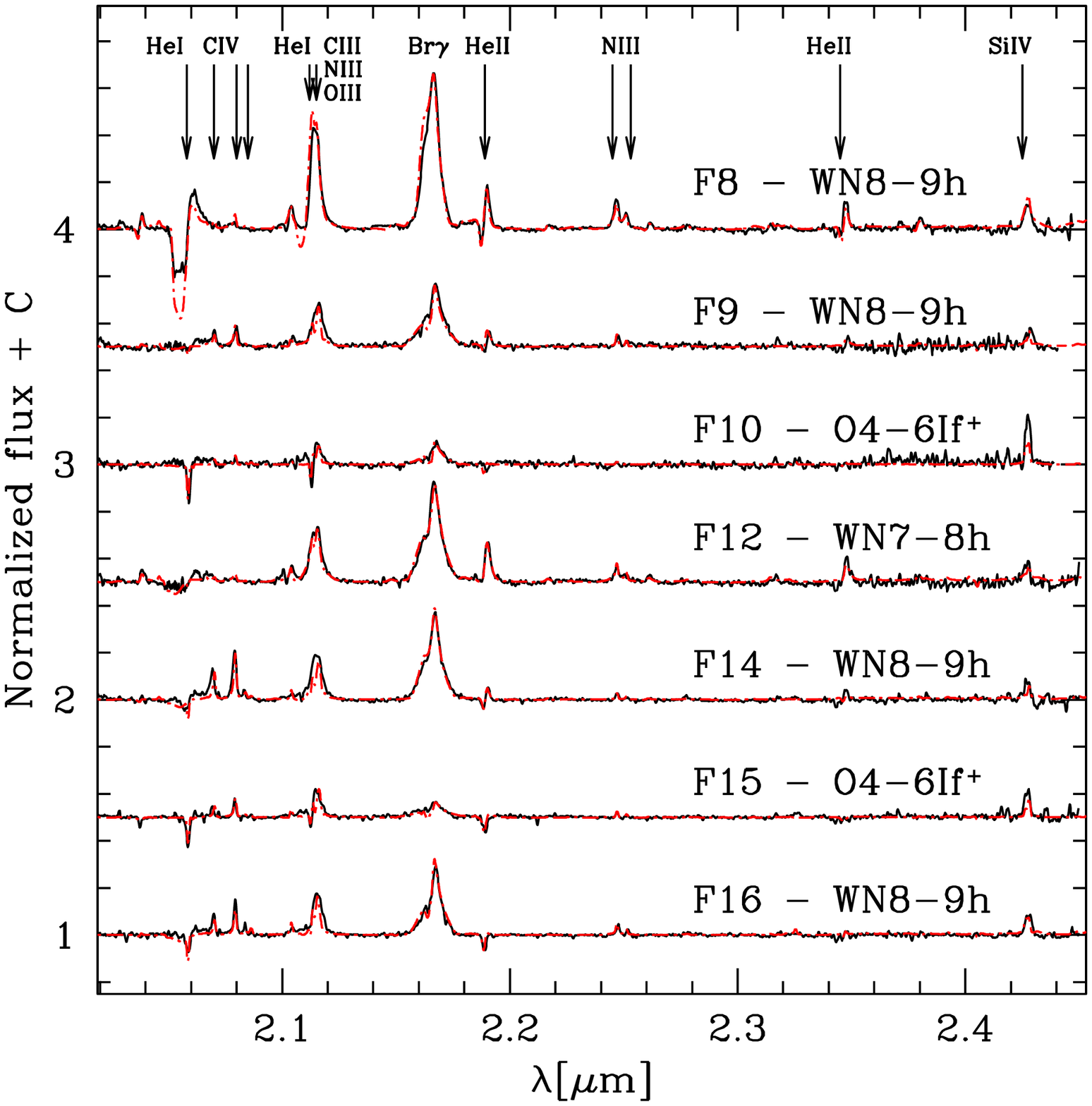}
\caption{Best fits (red dot-dashed lines) of the observed K-band spectra (black solid line). \label{fit_all_2}}
\end{figure*}

\begin{figure*}
\includegraphics[width=19cm,height=25cm]{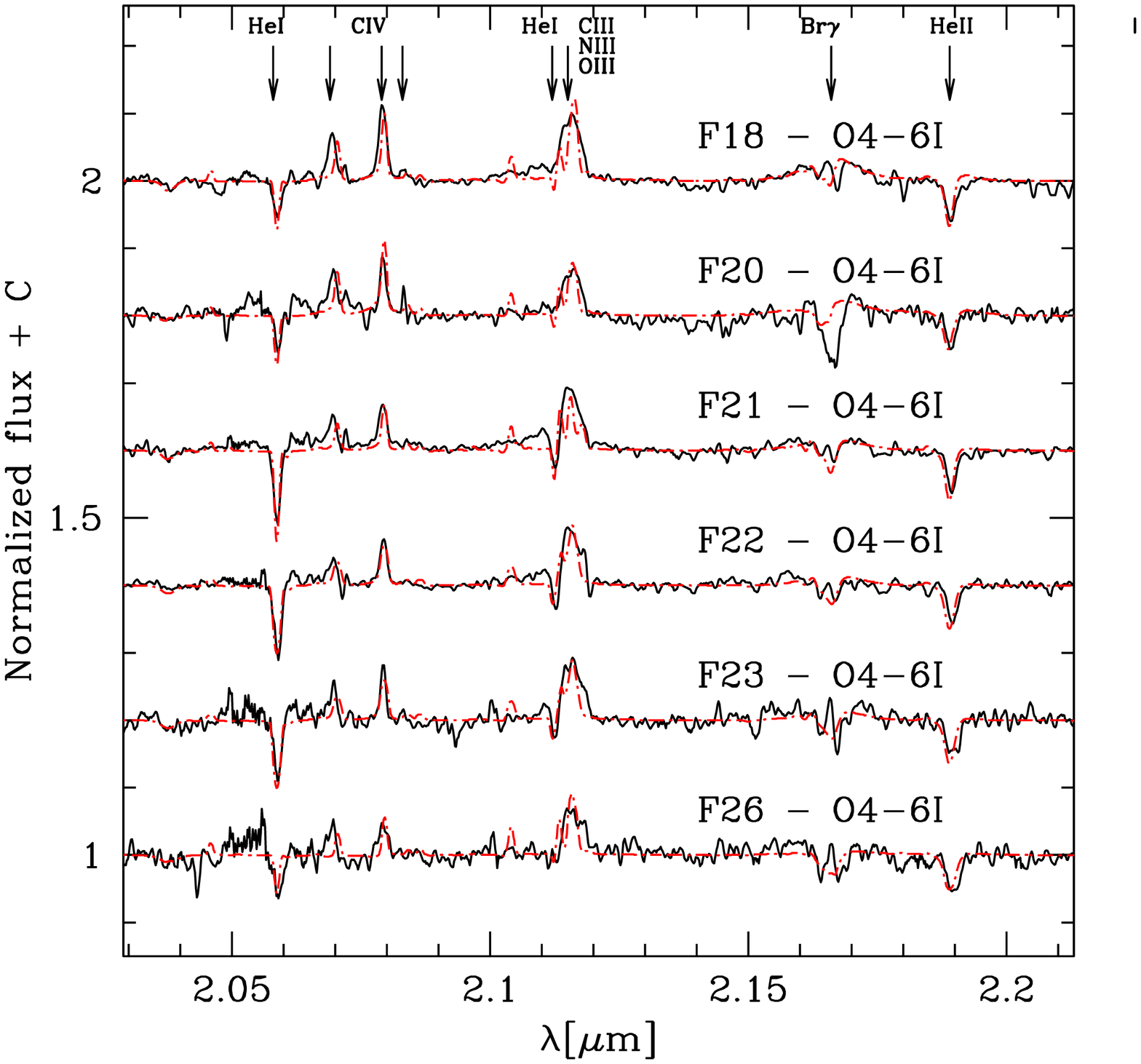}
\caption{Best fits (red dot-dashed lines) of the observed K-band spectra (black solid line). \label{fit_all_3}}
\end{figure*}

\begin{figure*}
\includegraphics[width=19cm,height=25cm]{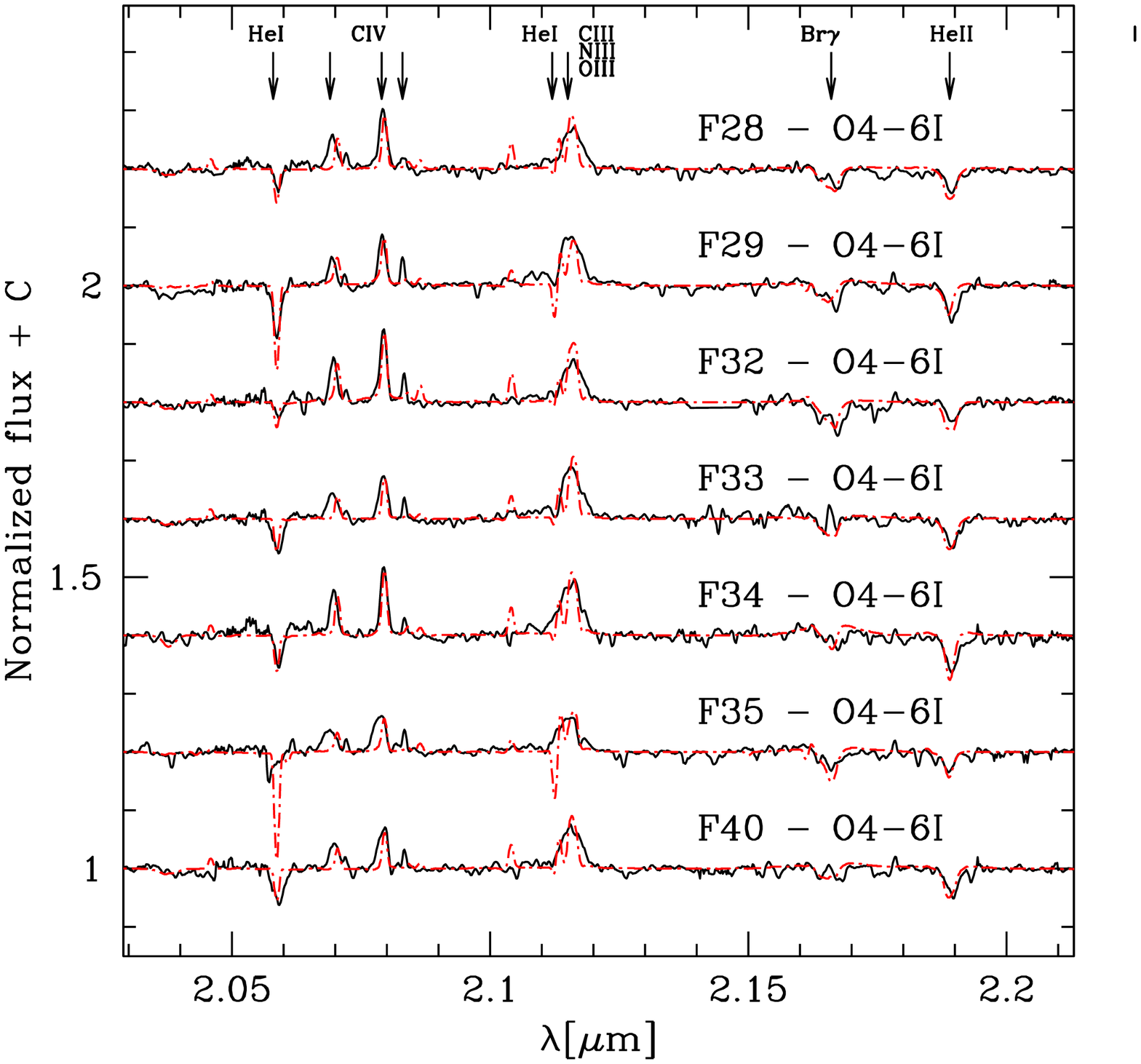}
\caption{Best fits (red dot-dashed lines) of the observed K-band spectra (black solid line). \label{fit_all_4}}
\end{figure*}

\end{appendix}

\end{document}